\documentclass[journal]{IEEEtran}
\IEEEoverridecommandlockouts

\usepackage{amsmath} 
\usepackage{amssymb}  
\usepackage{amsfonts} 
\usepackage{dsfont}

\usepackage[utf8]{inputenc}

\usepackage[pdftex]{graphicx}
\usepackage[usenames, dvipsnames]{color}
\usepackage{color,soul}
\setstcolor{red}

\usepackage[labelformat=simple]{subcaption}

\usepackage{multirow}
\usepackage{makecell}
\usepackage[linesnumbered,ruled]{algorithm2e}

\usepackage{array}
\usepackage{url}

\usepackage{nomencl}

\usepackage{amsthm}
\usepackage{tikz}

\theoremstyle{definition}

\theoremstyle{remark}

\usepackage{makecell}
\usepackage{amssymb}

\usepackage{lipsum}
\usepackage{multicol}

\usepackage{bm}

\usepackage{relsize}
\usepackage{hyperref}
\usepackage{cleveref}

\usepackage{epsfig}
\usepackage{graphicx}
\usepackage{array}
\usepackage{booktabs}
\usepackage{xcolor}
\usepackage{multirow}

\usepackage{textcomp}

\newcommand{\xhdr}[1]{\vspace{6pt} \noindent {\textbf{#1} }}

\newcommand{\abs}[1]{\left|#1\right|}

\usepackage{etoolbox}
\makeatletter
\patchcmd{\@makecaption}
  {\scshape}
  {}
  {}
  {}
\makeatother

\begin{document}
\title{\LARGE \bf
The Smart Black Box: A Value-Driven High-Bandwidth \\ Automotive Event Data Recorder} 

\author{Yu Yao and Ella M. Atkins
\thanks{This research has been supported by the National Science Foundation Award Number CNS 1544844.
The authors are with the Robotics Program, University of Michigan, Ann Arbor, MI, USA, 48109, {\tt\small \{brianyao,ematkins\}@umich.edu}}
}

\maketitle

\begin{abstract}
Autonomous vehicles require reliable and resilient sensor suites and ongoing validation through fleet-wide data collection. This paper proposes a Smart Black Box (SBB) to augment traditional low-bandwidth data logging with value-driven high-bandwidth data capture. The SBB caches short-term histories of data as buffers through a deterministic Mealy machine based on data value and similarity. Compression quality for each frame is determined by optimizing the trade-off between value and storage cost. With finite storage, prioritized data recording discards low-value buffers to make room for new data.
This paper formulates SBB compression decision making as a constrained multi-objective optimization problem with novel value metrics and filtering. The SBB has been evaluated on a traffic simulator which generates trajectories containing events of interest (EOIs) and corresponding first-person view videos. SBB compression efficiency is assessed by comparing storage requirements with different compression quality levels and event capture ratios. Performance is evaluated by comparing results with a traditional first-in-first-out (FIFO) recording scheme. Deep learning performance on images recorded at different compression levels is evaluated to illustrate the reproducibility of SBB recorded data.
\end{abstract}

\section{Introduction}
Autonomous vehicles (AVs) require verification and validation (V\&V) to minimize or eliminate the potential for incorrect perceptions, decisions, and actions.  The industry has used the Naturalistic Field Operation Test (NFOT) project to collect a large amount of driving data and has conducted Monte Carlo simulations to enable such V\&V~\cite{Zhao2017}.  
However, NFOT data indicate low exposure rates to EOIs~\cite{safety2015}, suggesting that a large amount of collected data  are of minimal to no interest thus could be discarded or logged with a very high compression loss factor. 

\begin{figure}[htbp]
    \centering
    \includegraphics[width=0.95\columnwidth]{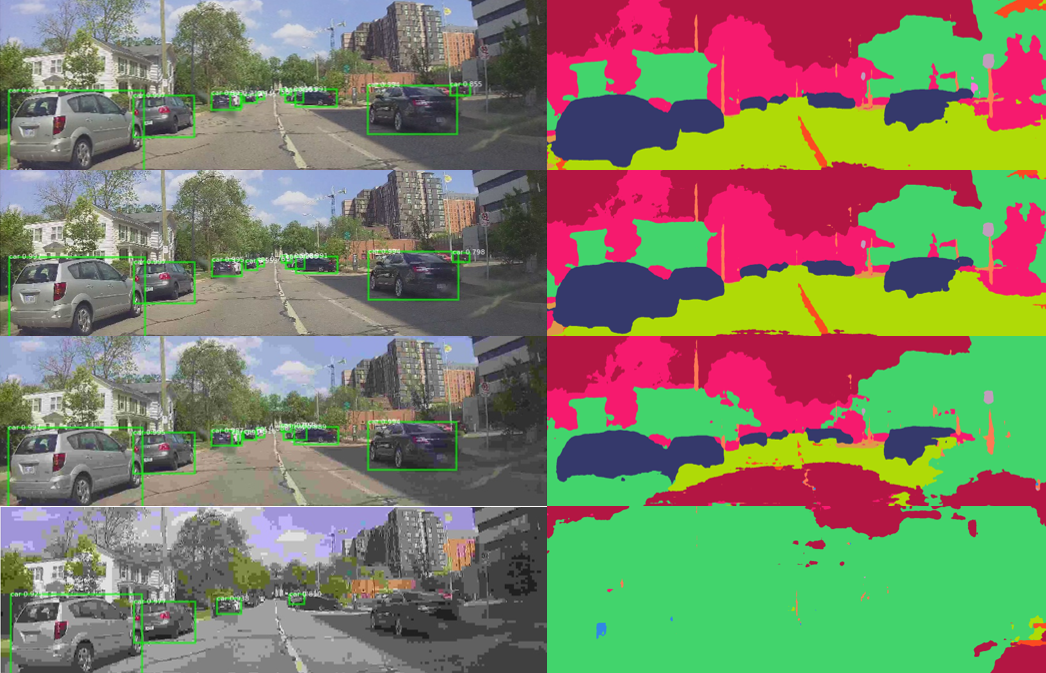}
    \caption{Object detection using Mask-RCNN~\cite{He2017} (left); semantic segmentation using DeepLabV2~\cite{CP2016Deeplab} (right).  From top to bottom the images are compressed with $1$, $0.5$, $0.1$ and $0.01$ quality by a JPEG algorithm.}
    \label{fig:compress_intro}
\end{figure}

Emerging AVs with redundant high-bandwidth sensors (e.g. camera, LiDAR) generate as much as $1$ GB/second of raw data, a figure that scales to $\sim2160$ TB/year given an average driving time per person of $660$ hours/year~\cite{forbes}. Effective capture of this raw data fleet-wide over the long-term is therefore challenging, motivating efficient data compression and discard capabilities. Recent advances in deep learning have motivated AV research in object detection~\cite{girshick2015fast,ren2015faster,He2017}, tracking~\cite{choi2015near,Xiang_2015_ICCV,wojke2017simple}, and semantic scene segmentation~\cite{girshick2014rich,long2015fully,badrinarayanan2015segnet,CP2016Deeplab}. However, these modules might fail when the raw data is recorded with significant compression.
Fig.~\ref{fig:compress_intro} illustrates how object detection and semantic segmentation are impacted by application of lossy compression (JPEG). Compression level (1, 0.5, 0.1, 0.01) represents an image quality parameter where $1$ means the highest quality.  These images show that with significant compression, learning algorithms cannot accurately reproduce results obtained in situ with raw image data even though the human eye can still succeed.   

This paper proposes a Smart Black Box (SBB) framework first introduced in~\cite{yao2018sbb} to record high-priority high-bandwidth raw data as a supplement to logging low-bandwidth processed data, e.g., from a Controller Area Network (CAN) bus. The SBB quantifies data value and optimizes the trade-off between stored data value and size. A decision indicates how much a given data frame should be compressed for recording. Buffers containing raw data are compressed and queued so that low-value long-term data are aged out when the SBB approaches its finite storage limit. 

Designing the targeted SBB functionality is challenging because there is no standard procedure for quantifying driving data value. Further, no metrics have been established to optimize long-term data collection for on-road vehicles. Additionally, globally optimizing data compression and deletion decisions would require buffering all data over a long-term trajectory. 

This paper proposes a \textit{data value information metric} based on events and data rarity to make locally-optimal compression decisions for each short-term buffer. Compressed data and their value are saved in a long-term priority queue to enable removal of the lowest-value data when finite storage limits are reached.
To evaluate SBB performance, we generated a  large-scale highway traffic dataset using a simulator that is capable of representing heterogeneous and interactive multi-vehicle traffic scenarios~\cite{Li2017game} and detected EOIs along each simulated driving trajectory. This paper studies four EOIs: lead car cut-in ($cutin$), host car hard-braking ($hard braking$), cut-in conflict ($conflict$) and $crash$. A frame with none of these EOIs is classified as a $normal$ frame or event. Values of these events are computed from their likelihood (rarity) among the dataset, and SBB compression and storage statistics are analyzed. This list of EOIs can be extended and generalized per the specific data collection task. 

This paper offers four primary contributions. First, we formulate a value-based data recorder, the SBB, to store high-bandwidth long-term driving data based on data value metrics. This is the first value-based automotive event data recorder to our knowledge.  Second, we propose a deterministic Mealy machine (DMM) to track incoming data by value and similarity to enable high-value data and data in a common context to be buffered together. Third, we define a multi-objective constrained optimization strategy to define lossy compression factor for each buffered data frame based on computed data value for the current frame, data value for surrounding frames, and storage cost. Finally, we propose a simple but effective strategy for managing finite onboard storage to ensure the highest-value data can be saved over the long-term.
To demonstrate the impact of lossy compression ratio on  SBB recorded data, we test popular deep learning models for object detection and semantic segmentation on first-person view images from a high-fidelity simulator called The Open Racing Car Simulator (TORCS)~\cite{TORCS}. We show that images compressed by the SBB have smaller size but can still reproduce events of interest with deep learning image processing.  

Note that the SBB formulation were first introduced in~\cite{yao2018sbb}; this manuscript substantially extends the previous paper with respect to methods, metrics, case studies, and results.

This paper is structured as follows. Following background (Section \ref{sec:background}), Section \ref{sec:design} introduces the SBB framework and presents key definitions. Section \ref{sec:traffic_eoi} defines driving data and EOIs used in this paper, followed by data value and similarity metrics specifications in Section \ref{sec:value_similarity}.
Section \ref{sec:LBO} describes the optimization strategy used along with our value-based algorithm for finite long-term data storage  management.  Section \ref{sec:case} presents results from a simulation case study followed by a brief conclusion in Section \ref{sec:conclusion}.

\section{Related Work}\label{sec:background}
\subsection{Event Data Recorder}

The automobile event data recorder (EDR) was developed by manufacturers to analyze precursor and crash data with impact-triggered recording ~\cite{dasilva2008analysis, gabler2004crash}. The EDR captures low-bandwidth data including but not limited to vehicle speed, engine speed, throttle, brake status, and acceleration. 
Improvements on EDRs have extended the event list an EDR can detect~\cite{Takeda2012} and support system execution replay~\cite{Narayanasamy}. An in-vehicle data recorder was proposed in~\cite{Perez2010} to record heterogeneous data from asynchronous onboard sensors.
However, these publications do not address the trade-off between data compression losses and finite storage constraints that exist in long-term high-bandwidth data collection.

The SBB is designed to operate long-term to record high-bandwidth data without human intervention. An essential challenge is to selectively remove recorded data as needed to make room for new data.
Two models have been adopted in previous data recorders. The most common data recorder continuously writes data until the storage is full then terminates the recording. The newest data is dismissed in this model. Some EDR systems use circular buffers to record data so that once the storage is full, the oldest data is aged out with a first-in-first-out (FIFO) queue. This model values new data over old data without processing data contents.  

In~\cite{ganesan2003evaluation}, multi-resolution storage was supported by generating coarse representations of raw data, called summaries. Summaries in the database are then aged out by a user-defined aging function. Such an approach was applied in~\cite{diao2007rethinking} where the elimination process considered the frequency at which data is queried. This method focuses on database maintenance and query instead of on-line data collection. 
In this paper, we age out data based on data value metrics and storage limitation. The optimization of compression quality is solved and the reproducibility of computer vision algorithms on compressed data is analyzed. Because the SBB is designed for long-term data collection, query strategies are not studied in our work.

\subsection{Event Detection for AVs}

Certain EOIs can be straightforwardly detected based on pre-defined rules or safety envelopes such as sudden deceleration or insufficient following distance~\cite{Zhao2017, Takeda2012}.
Other events are detected from driver behavior recognition or driving environment characterization.
Driver behavior has been assessed to-date by training statistical models or feature extraction models based on CAN bus signals~\cite{taylor2016anomaly, Liu2017, Miyajima2016} and driver observation camera data frames~\cite{li2016driver}. Most environment detection research focuses on recognizing behaviors in surrounding vehicles~\cite{Sivaraman2013, Lefvre2014, Shirazi2015} and/or pedestrians~\cite{Popoola2012, Schulz2015, Yi2016}, after which an EOI analysis of human-vehicle interaction is possible. The detection of road conditions such as potholes has also been investigated in~\cite{li2016optimal}.

Although many event and anomaly detection techniques have been designed for automated driving and risk recovery~\cite{taylor2016anomaly,Lampiri2017}, few are used to guide data collection and compression. 

\subsection{Public Driving Datasets for Autonomous Vehicles (AV)}

Emerging autonomous vehicle research has led to several driving data collection projects. NFOT datasets/databases and vision oriented driving data sets are particularly relevant to this paper thus are summarized below. The proposed SBB offers a novel data collection pipeline to mitigate drawbacks of these datasets by collecting high-quality, high-interest, high-bandwidth data over an extended time period.

\subsubsection{NFOT database}
Researchers undertake NFOT projects to collect a large amount of naturalistic driving data to support V\&V efforts~\cite{ervin2005automotive, leblanc2006road, victor2010sweden, sayer2011integrated}. To-date NFOT data has mainly been used for driver modeling~\cite{wang2017driving}, safety assessment~\cite{Zhao2017} and connected vehicle research. Data acquisition systems (DAS) are distributed among fleets of vehicles to collect data on roads. 

A 100-car naturalistic dataset~\cite{neale2005overview} includes approximately $2,000,000$ miles, almost $43,000$ hours of video, electronic driver, and vehicle performance data collected over an approximately one year period. Based on this data set, an event database has been constructed with crashes, near crashes and other incidents labelled manually. The video stream, the major high-bandwidth data in this dataset, has been compressed by the MPEG-1 algorithm, resulting in a downsampled low-resolution video at $\sim7.5\,Hz$.

A more recent NFOT dataset is the Safety Pilot Model Deployment (SPMD)~\cite{bezzina2014safety} which contains approximately $ 1.7$ million miles of driving data with nearly $64,400$ hours of collection over an approximately two year period. A $656\,GB$ database containing  numerical (low-bandwidth) data has been constructed from the SPMD data set and indexed by driving scenarios such as cut-in and lane-change~\cite{zhao2017trafficnet}. There are $17\,TB$ of video data recorded from four grey-level cameras, compressed with the MPEG-4 algorithm and downsampled to $10\,Hz$ (forward and cabin views) and $2\,Hz$ (side views). 

These NFOT projects take advantage of database structures to organize large-scale data and utilize compression techniques to record high-bandwidth videos. However, recorded videos are mainly used by human evaluators to find causations of events. Automatic scene recognition and reconstruction have not yet been applied due to the poor quality of compressed video streams. Also, video data lost can become an issue, as in~\cite{bezzina2014safety} that discusses the possible loss of valuable data after DAS storage is filled to capacity after a long period of driving. 

\subsubsection{Vision oriented driving data sets}

One of the most significant topics in the autonomous driving field is environment perception including but not limited to object detection, tracking, semantic scene segmentation and localization. Many vision-oriented datasets have been collected to support benchmark testing, such as KITTI~\cite{Geiger2013IJRR}, Cityscape~\cite{Cordts2016Cityscapes} and BDD100K~\cite{yu2018bdd100k}. These data sets contain high-quality high-bandwidth data (images, videos, and LiDAR point clouds) that can be used for validating autonomous driving algorithms.

KITTI is a relatively small data set containing only $7,481$ labeled images ($12\,GB$) with corresponding $29\,GB$ point cloud data. Cityscape~\cite{Cordts2016Cityscapes} is similar to KITTI but with larger scale ($24,999$ labeled images for $55\,GB$) which mainly serves  as a benchmark for semantic scene understanding. BDD100K contains $100,000$ HD short video clips ($1.8\,TB$) collected over $1,100$ driving hours with different times, weather conditions and driving scenarios. 

While these data sets offer good-quality data capture, they are still too small to serve for comprehensive autonomous vehicle V\&V. Also, none of them contains appreciable or sufficient EOIs that challenge self-driving algorithms. This motivates our SBB which directly addresses this challenge by offering a way to collect high-quality, high-interest, high-bandwidth data long-term over a vehicle fleet by prioritizing EOI data capture.  The Smart Black Box (SBB) proposed in this paper utilizes event analysis and lossy compression techniques to determine which data frames should be recorded together,  what compression factor should be applied to each frame, and what data frames should be discarded to meet onboard storage constraints.

\section{Smart Black Box Design}\label{sec:design}

This paper proposes a generalized SBB framework to realize efficient data collection from emerging high-bandwidth sensors such as LiDAR and cameras given finite local storage. The SBB minimizes the size of recorded data and maximizes recorded data value by determining how much each data frame should be compressed. The following questions are addressed in this work:

\begin{itemize}
    \item How are data values quantified? 
    \item What compression factor should be applied to each data frame to trade off data value and data storage size?
    \item How does the SBB select data to discard given finite storage constraints?
    \item How do we quantify or evaluate data recording performance, i.e., what metrics should be applied?
\end{itemize}

Three data storage stages are implemented in the SBB: buffer, long-term storage and cloud database. Buffers are used to temporally cache seconds or minutes of raw data in real-time. Long-term storage relies on a finite onboard storage device capable of recording data collected over days or weeks given normal usage. The cloud database stores and manages data from vehicle fleets over months and years; data is retrieved later for post-processing.

\begin{figure}[htbp]
    \centering
    \includegraphics[width=1.0\columnwidth]{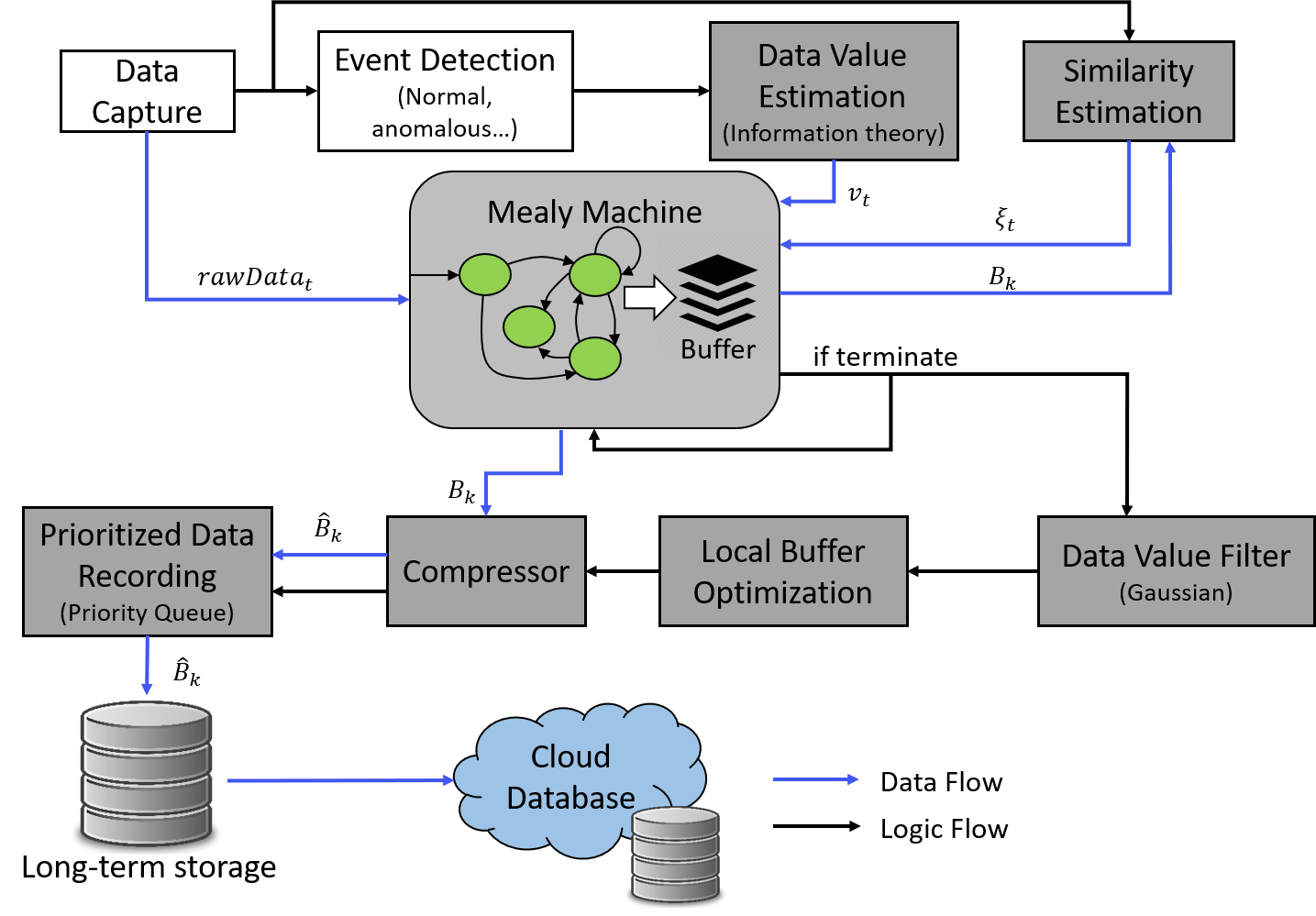}
    \caption{Flow chart of the SBB data recording process. Gray blocks represent SBB functions; blue blocks represent data storage and monitoring. Black arrows show the logic flow while blue arrows show the data flow.}
    \label{fig:SBB_FSM}
\end{figure}

SBB functionality is proposed in Fig.~\ref{fig:SBB_FSM}. At each time step one data frame is collected. Each frame is classified based on event detectors; a scalar in $[0,1]$ is computed as frame data value. Similarity between a new frame and adjacent buffered frames is also computed as a scalar in $[0,1]$. A DMM is applied to automatically manage the data buffering process. The inputs to the DMM are the data value, similarity and buffer size. It outputs buffer operation instructions, e.g.,  writing to buffer and emptying a buffer. The DMM formulation is detailed in Section \ref{sec:DMM}. Once the DMM terminates, an optimization problem is solved over the buffered data to determine optimal compression quality for each frame. This process is called local buffer optimization (LBO). A data value filter can be applied to smooth the estimated value. Buffered raw data are compressed and recorded in long-term storage and are sorted based on their values. Once onboard data storage is filled, the lowest-value data are discarded to make room for new high-value data (i.e. prioritized data recording). Given internet access, stored data can be uploaded to a cloud then removed from local storage. Note that data uploading and cloud database management are not studied in this paper to focus attention on compression and discard decision-making.

The SBB offers several advantages. First, DMM buffer tracking enables high-value data and data with a common context to be buffered together.  Data value filtering and local buffer optimization (LBO) ensures contextual frames of EOIs are considered. Second, by separating LBO and long-term storage prioritization, the SBB makes locally optimal compression decisions which reduces memory and time complexity relative to long-term (global) optimization.  Third, a long-term data storage prioritization scheme enables rapid identification of the lowest-value data to facilitate deletion as needed. Note that conventional database management methods are not applied here since the local storage is designed for data collection with no requirement for high-speed retrieval.

Some key definitions and mathematical notations used in Fig.~\ref{fig:SBB_FSM} are introduced below:

\begin{itemize}
    \item \textbf{Long-term storage size}: The maximum local storage capacity (e.g. in MegaBytes) that the SBB can utilize, denoted $M$. 
    \item \textbf{Frame}: The sensor data received at each time as well as its value $v_t$ and storage cost (size) $c_t$, represented as $f_t=(v_t,c_t,rawData_t)$.
    \item \textbf{Decision}: $d_t\in[0,1]$ is the desired quality to compress $rawData_t$, $0$ and $1$ denote the lowest and highest data qualities, respectively.
    \item \textbf{Local buffer}:  Short-term sequential frames are cached in a local buffer $B_k$, where $k$ is buffer index. There is $f_t\in B_k$ if a frame $f_t$ is cached in buffer $B_k$. Buffer length is a scalar $|B_k|$ determined by the DMM. 
    \item \textbf{Recorded buffer}: Compressing the local buffer based on LBO output yields recorded buffer $\hat{B}_k$. Each recorded buffer is saved in the database and can be retrieved by two key parameters: the temporal index $k$ and/or the flagged event type of the buffered data. Definition of event types is introduced later. 
    \item \textbf{Local buffer decision vector}: Decisions for every frame in $B_k$ form the decision vector $D_k$. 
\end{itemize}

\section{Traffic Data and Events of Interest}\label{sec:traffic_eoi}
\begin{figure}[htbp]
    \centering
    \includegraphics[width=0.7\columnwidth]{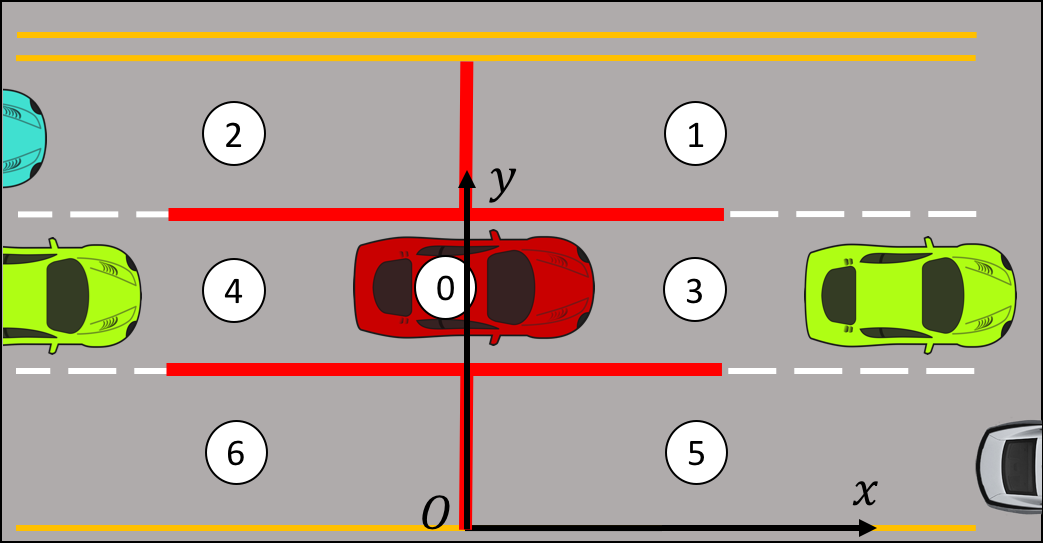}
    \caption{Three-lane traffic scenario and reference frame. Circled numbers indicate the host vehicle (in red) and closest vehicles in six surrounding regions, separated by red lines.}
    \label{fig:traffic_frame}
\end{figure}

This paper focuses on SBB data collection in multi-lane highway traffic scenarios where one host vehicle and multiple participant vehicles are present (see Fig. \ref{fig:traffic_frame}). This section introduces a simplified data representation and EOIs that can be detected from this traffic scenario. A more complex and comprehensive set of EOIs would be developed over a long-term for full deployment of the generalized SBB framework in future work.

\subsection{Data Representation}

The proposed data reference frame is depicted in Fig. \ref{fig:traffic_frame}, where the origin $O$ is the projection of host vehicle centroid on the road right edge. We assume full observability of host and all nearby vehicle locations ($x,y$) and speeds ($\dot{x}$) from processed sensor data (e.g. CAN Bus, LiDAR, radar,  camera). Other physical parameters are ignored and the lane widths and locations are assumed constant for simplicity. In this paper we only consider the closest vehicles in six regions: front left (1), rear left (2), front center (3), rear center (4), front right (5), rear right (6). Also, we compute $x_1$ to $x_n$ as relative distance to the host car so that $x_0=0$ can be ignored, resulting in a $20$ dimensional feature vector: 

$$X = [y_0,\dot{x}_0,x_1,y_1,\dot{x}_1,...,x_6,y_6,\dot{x}_6]$$

\noindent where subscript $0$ indicates host vehicle features and $1...6$ are the six neighbor cars. For a region where no vehicle exists, we set $x_i=100\,m$, $\dot{x}_i=0\, m/s$ and $y_i$ equal to the location of the corresponding lane center line.

\subsection{Events of Interest (EOIs)}
We classify each observed data frame as either $normal$ or one of the four EOIs detailed in~\cite{yao2018sbb}: $cutin$,  $hard braking$, $conflict$ and $crash$. A $normal$ frame is a frame that is not classified as any of the EOIs. Detection of the four EOIs are based on pre-defined physical thresholds as summarized below:

\subsubsection{$Cutin$} A $cutin$ event is recognized when the closest front vehicle, represented by $(x_i,y_i,\dot{x}_i)$, enters the lane of the host vehicle as shown in Fig.~\ref{fig:cutin}. The cut-in range is $R=x_i-x_0$.
Let $\dot{y}_i$ be the lateral velocity of the cut-in vehicle, and $w_{ln}$ and $w_c$ be the widths of lane and vehicle, respectively. A cut-in event is defined by

\begin{equation}\label{eq:cutin}
    \begin{split}
                    0<y_i-y_0<\frac{w_{\text{ln}}+w_\text{c}}{2} \text{ and } \dot{y}_i < 0 \\
        \text{or } -\frac{w_{\text{ln}}+w_\text{c}}{2}<y_i-y_0<0 \text{ and } \dot{y}_i > 0 
    \end{split}
\end{equation}

\subsubsection{$Hardbraking$} 

A $hardbraking$ event occurs when the deceleration of a car is greater than a hard deceleration threshold $\ddot{x}_{\text{hb}}$ as in Fig.~\ref{fig:breaking}. In this paper we define $\ddot{x}_{\text{hb}}\approx-4.4\,m/s^2$ per~\cite{Takeda2012}.

\subsubsection{$Conflict$} A $conflict$ event is when the host car is in the proximity zone of the lead car during the cut-in event as shown in Fig.\ref{fig:conflict}. The proximity zone of a lead car is the rectangle area bounds its geometric contour from 4 feet in front of its front bumper to 30 feet behind its rear bumper~\cite{lanechange2004lee}. Its length and width are defined as $(l_{\text{pr}}, w_{\text{c}})$. 

\subsubsection{$Crash$} 
\begin{figure}[htbp]
    \centering
    \begin{subfigure}[t]{0.45\columnwidth}
        \centering
        \includegraphics[width = 0.9\columnwidth]{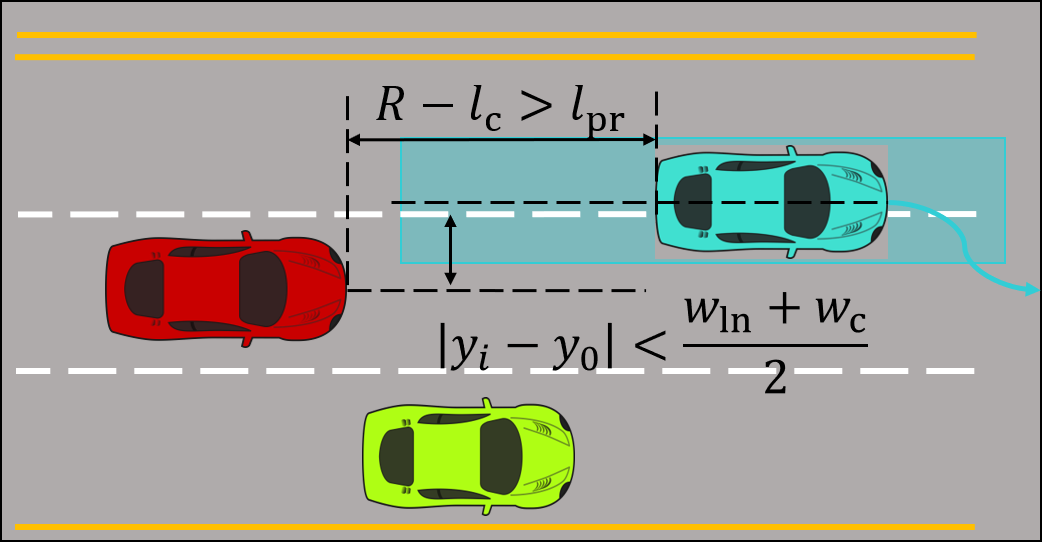}
        \caption{Cutin}
        \label{fig:cutin}
    \end{subfigure}
    ~
    \begin{subfigure}[t]{0.45\columnwidth}
        \centering
        \includegraphics[width = 0.9\columnwidth]{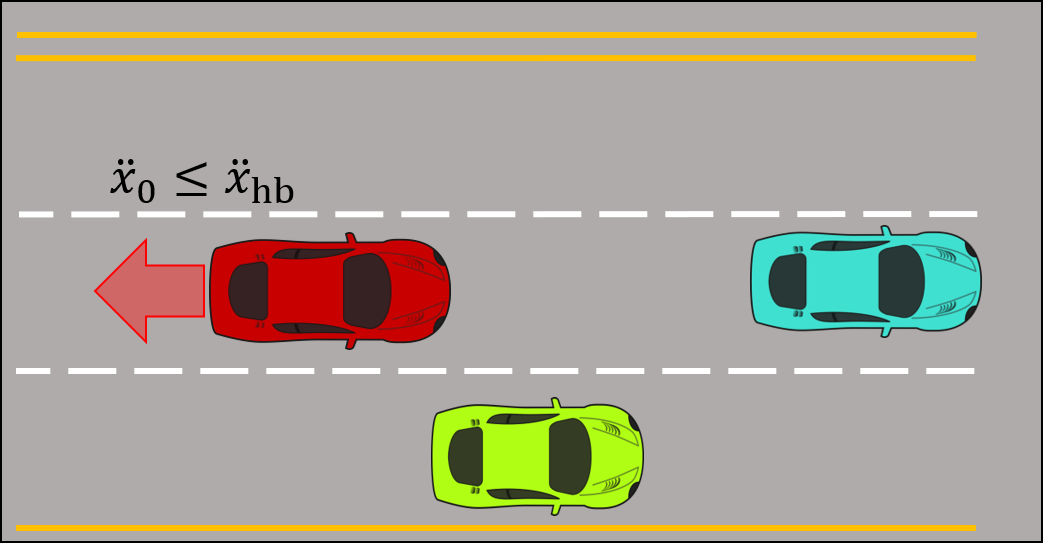}
        \caption{Hardbraking}
        \label{fig:breaking}
    \end{subfigure}
    
    \begin{subfigure}[t]{0.45\columnwidth}
        \centering
        \includegraphics[width = 0.9\columnwidth]{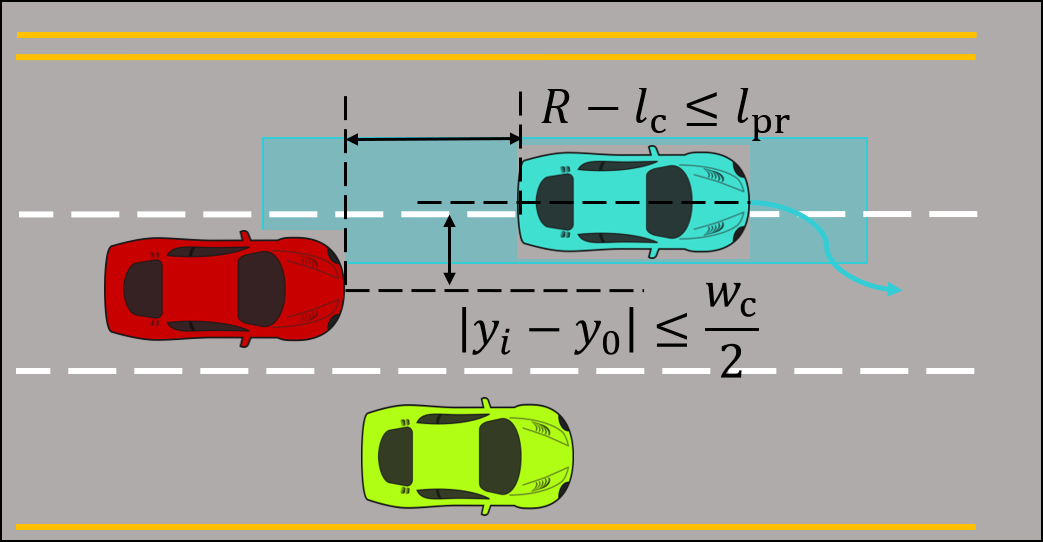}
        \caption{Conflict}
        \label{fig:conflict}
    \end{subfigure}
    ~
    \begin{subfigure}[t]{0.45\columnwidth}
        \centering
        \includegraphics[width = 0.9\columnwidth]{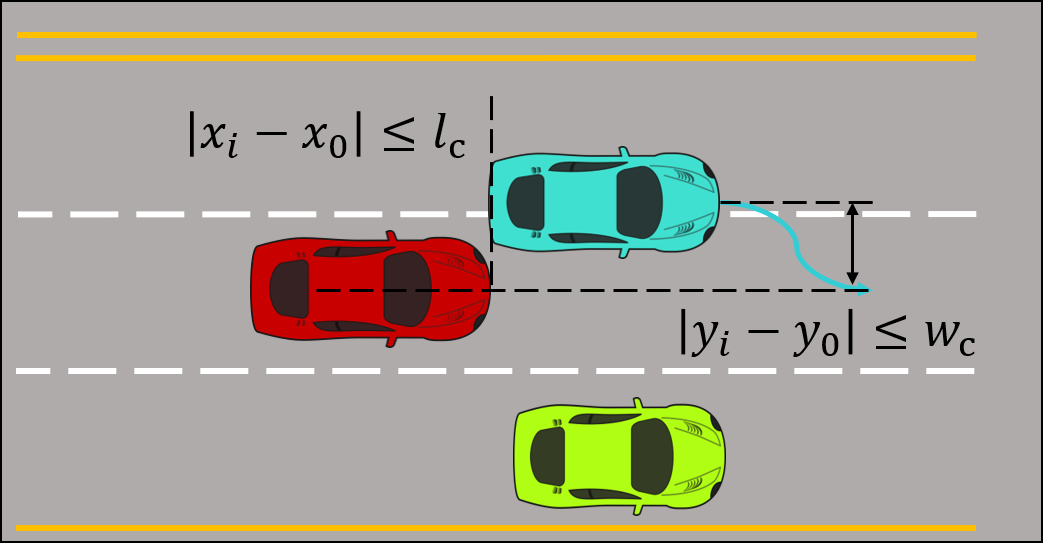}
        \caption{Crash}
        \label{fig:crash}
    \end{subfigure}
    \caption{Four pre-defined EOIs. The blue rectangle indicates proximity zone of the blue car (better in color). }
    \label{fig:events}
\end{figure}

A $crash$ event occurs when one car collides with another car from any direction. Since car yaw angle is ignored for simplicity, we detect a crash by 
\begin{equation}
    \abs{x_i-x_0} \leq l_{\text{c}} \text{ and } \abs{y_i-y_0} \leq w_{\text{c}}
\end{equation}
where $l_{\text{c}}$ is vehicle length. Fig.~\ref{fig:crash} shows a crash event.

Although multiple EOIs may be detected in a single frame, we mark each frame with the single highest-value EOI to to simplify value assignment. Therefore the $normal$ event plus the four EOIs constituted the event space $\mathbb{E}$ for data value assignment, defined as:
\begin{align*}
    \mathbb{E} & =\{\varepsilon_1,\varepsilon_2,\varepsilon_3,\varepsilon_4,\varepsilon_5\} \\
    & = \{normal, cut in, hard braking, conflict, crash\}
\end{align*}

The above list of EOIs can be extended and generalized for any data collection task. Advanced detection models can be applied to detect more complicated EOIs. In the following section we present a generalized value metrics computation which can be applied to any EOI as long as an event likelihood probability is provided.

\section{Data Value and Similarity Metrics}\label{sec:value_similarity}

This section introduces our data value and similarity metrics based on the event definition from Section \ref{sec:traffic_eoi}. Data value and similarity will later be applied by the DMM for buffer definition and data tracking.

\subsection{Data Value Metrics}

We assume a large naturalistic driving data set which contains all previously defined EOIs has been collected and processed. Given a frame $f_t$ whose corresponding event type is $\varepsilon_j\in\mathbb{E}$, data value is based on the likelihood (rarity) of $\varepsilon_j$.
According to information theory, a lower-probability event carries more information than a higher-probability event, so
$v_t=v(\varepsilon_j)$ can be estimated as the information measure of $\varepsilon_j$.
We use previously defined EOIs and assume $100\%$ detection confidence for simplicity. Value estimates of different events are given in this section.

\subsubsection{Constant value events} We assume a $normal$ event has constant low value while $hardbraking$, $conflict$ and $crash$ events have constant high values. Values of these events $\varepsilon_j\in\{\varepsilon_1,\varepsilon_3,\varepsilon_4,\varepsilon_5\}$ are computed using \eqref{eq:eoi_value} given $\Pr(\varepsilon_j)$ the event likelihood. 

\begin{equation}\label{eq:eoi_value}
    v(\varepsilon_j) = -\log_2\big(\Pr(\varepsilon_j)\big)
\end{equation}

In this paper we set $v(\varepsilon_5)=1$ (highest value) and the values of other events are normalized over $[0,1]$.

\subsubsection{Dynamic value event}

The value of a $cutin$ event is not a constant but a function of cut-in range $R$. 
Given a $cutin$ event, the conditional probability density function (PDF) of $R$ is represented by $\Pr(R|\varepsilon_2)$. Large $R$ indicates a low-value $cutin$, which is observed in the majority of the dataset. Overly small $R$ (e.g. $15\, m$) and overly large $R$ (e.g. $100\,m$) are rare in naturalistic driving data, but only small $R$ contains high value. Thus the value of a $cutin$ event with a measured $R_t$ can be computed from \eqref{eq:cutin_value}: 

\begin{equation}\label{eq:cutin_value}
    v(R_t,\varepsilon_2)  =-\log_2\big(\Pr(R<R_t|\varepsilon_2)\,\Pr(\varepsilon_2)\big)
\end{equation}
where $\Pr(\varepsilon_2)$ is the probability of $cutin$ events computed from the dataset. 

\begin{figure}[htbp]
    \centering
    \begin{subfigure}[t]{0.9\columnwidth}
        \centering
        \includegraphics[width = 0.9\columnwidth]{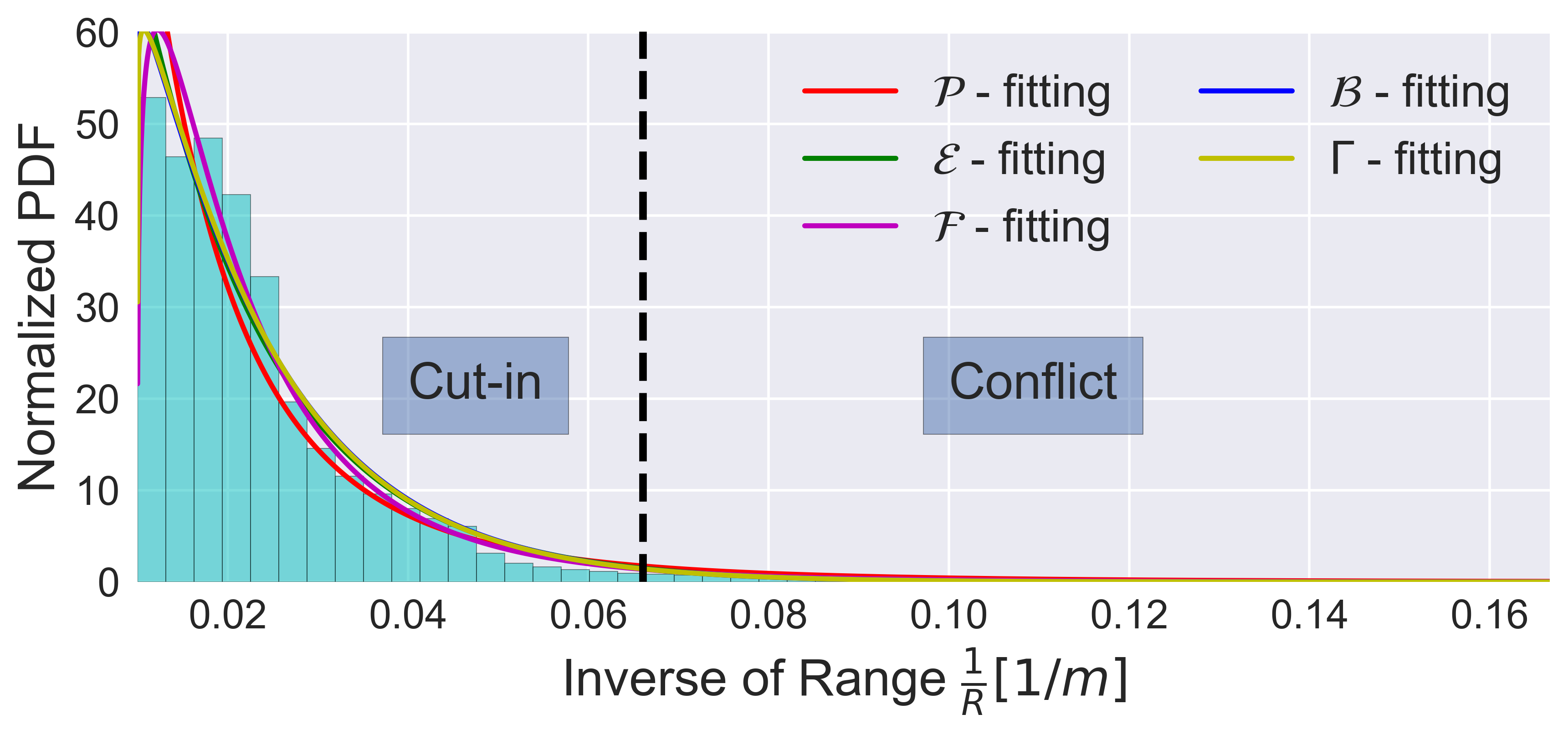}
        \caption{}
        \label{fig:R_inv}
    \end{subfigure}
    
    \begin{subfigure}[t]{0.9\columnwidth}
        \centering
        \includegraphics[width = 0.9\columnwidth]{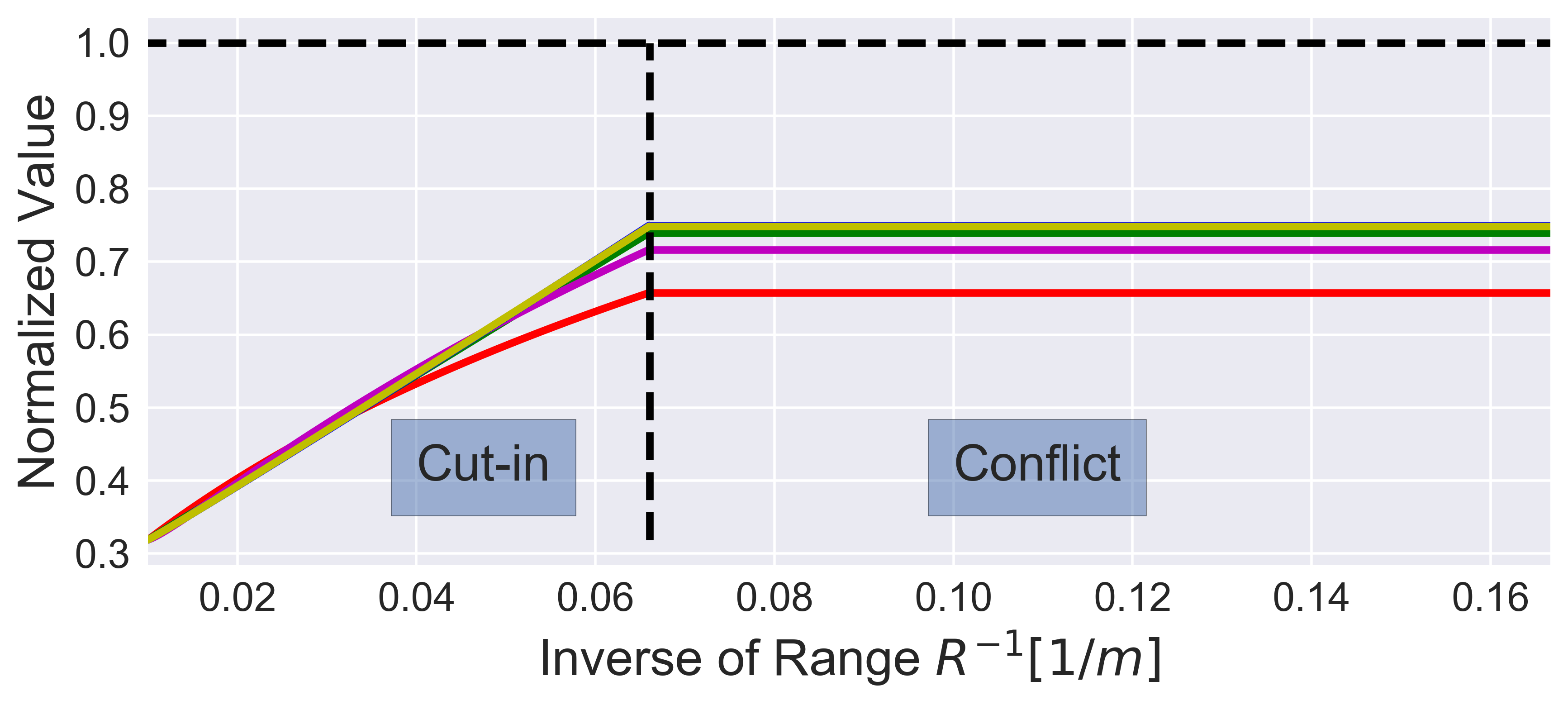}
        \caption{}
        \label{fig:cutin_value}
    \end{subfigure}
    \caption{Conditional PDF of inverse cut-in range $\Pr(R^{-1}|\varepsilon_2)$ and the calculated data value $v(R_t,\varepsilon_2)$ (normalized)}
    \label{fig:cutin_dist_value}
\end{figure}

In this paper, we use the conditional PDF of $R^{-1}$ instead of $R$ as suggested in~\cite{Zhao2017} to put the small $R$ value in the tail of the distribution. Fig.\ref{fig:cutin_dist_value} shows the fitting result of $\Pr(R^{-1}|\varepsilon_2)$ with Pareto distribution ($\mathcal{P}$), exponential distribution ($\mathcal{E}$), f distribution ($\mathcal{F}$), beta distribution ($\mathcal{B}$), and gamma distribution ($\Gamma$)~\cite{Zhao2017,wang2017driving}.

We use the Bayesian Information Criterion (BIC) for distribution fitting model selection since all candidate models are in the exponential family~\cite{Wit2012}. The BIC is computed as 

\begin{equation}
    BIC(\Theta, \mathcal{M}) = k\ln{n} - 2\sum_{i=1}^{n}\ln{\Pr\big(R_t^{-1}|\varepsilon_2;\Theta,\mathcal{M}\big)}
\end{equation}
where $n$ is the number of samples, $\mathcal{M}\in\{\mathcal{P},\mathcal{E},\mathcal{F},\mathcal{B},\Gamma\}$ is a candidate model, and $\Theta$ is the parameter vector with length $k$ that maximizes the likelihood. The model with lowest BIC is selected which for this paper is the $\mathcal{F}$ distribution with $\Theta=[\theta_1, \theta_2]$. The fitted conditional PDF is given in \eqref{eq:cutin_pdf}.

\begin{equation}\label{eq:cutin_pdf}
    \begin{split}
    & \Pr(R^{-1}|\varepsilon_2) = \Pr(R^{-1}|\varepsilon_2;\Theta,\mathcal{F}) \\ 
    & = \frac{1}{\beta(\frac{\theta_1}{2},\frac{\theta_2}{2})} (\frac{\theta_1}{\theta_1})^{\frac{\theta_2}{2}}R^{1-\frac{\theta_1}{2}}(1+\frac{\theta_1}{\theta_2}R^{-1})^{-\frac{\theta_1+\theta_2}{2}}
    \end{split}
\end{equation}

where $\beta(a,b) = \frac{(a-1)!(b-1)!}{(a+b-1)!}$. Equation \eqref{eq:cutin_value} can be written as \eqref{eq:cutin_value_inv} using the fitted distribution of $R^{-1}$.

\begin{equation}\label{eq:cutin_value_inv}
    \begin{split}
         & v(R_t,\varepsilon_2)\\
         = & -\log_2\Big[\Big(1-\int_{0}^{R_t^{-1}} \Pr(R^{-1}|\varepsilon_2)\,dR^{-1}\Big)\,\Pr(\varepsilon_2)\Big]
    \end{split}
\end{equation}

\subsection{Data Similarity Metrics}
We compute a similarity metric $\xi_t(f_t, B_k)$ to represent the similarity of a new coming data frame $f_t$ with the current buffer $B_k$. A high similarity between $B_k$ and $f_t$ indicates that they may belong to the same driving scenario so that $f_t$ might be appended to $B_k$ for completeness. This similarity metric together with the data value metric are input to the DMM to determine whether to buffer $f_t$ together with $B_k$ or not, as introduced in Section~\ref{sec:DMM}.

Consider the current buffered data time series $B_k$ with sequential feature vectors 
$[X_1,X_2,...,X_{|B_k|}]$ and a new single frame $f_t$ with feature vector $X_t$. 
The difference between $B_k$ and $f_t$, $\Delta(f_t,B_k)$, is defined as the standardized Euclidean distance between $X_t$ and previous feature vectors in \eqref{eq:difference}. Note that all $X$ are normalized to $[0,1]$.
\begin{equation}\label{eq:difference}
    \Delta(f_t, B_k) = \sqrt{\mathlarger{\sum}\limits_{j=1}^{20}\frac{(X_{t}^{(j)} - \mu_j)^2}{\sigma_j^2}}
\end{equation}
where $X_{t}^{(j)}$ is the $j$th feature of the new feature vector, and $\mu_j$ and $\sigma_j$ are the mean and standard deviation of the $j$th feature in $B_k$, respectively. The similarity score $\xi\in(0,1]$ is computed in \eqref{eq:similarity}. The higher the $\xi$ value, the more similar $f_t$ is to $B$.
\begin{equation}\label{eq:similarity}
    \xi(f_t,B_k) = e^{-\Delta(f_t,B_k)}
\end{equation}

\section{Online Data Buffering with a Deterministic Mealy Machine}\label{sec:DMM}

\begin{figure}[htbp]
    \centering
    \includegraphics[width=1.0\columnwidth]{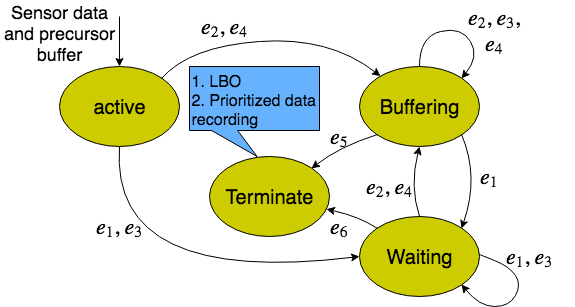}
    \caption{DMM for data buffer tracking decisions. The blue box highlights SBB actions executed when each buffer tracking DMM execution sequence terminates.}
    \label{fig:state_machine}
\end{figure}

The SBB must decide when to start buffering data and when to stop and send the data to the LBO module, i.e., the start and end points of data segments. This paper denotes this decision \textit{buffer tracking}. The buffer tracking process is modeled as a deterministic Mealy machine (DMM)~\cite{2008DES} as shown in Fig.\ref{fig:state_machine}. The DMM is defined as a 6-tuple $(S, S_0,\Sigma,\Lambda,T,G)$; each element is introduced below.

\begin{itemize}
    \item States: $S = \{s_i\}_{i=1}^4 = $ \textit{\{active, buffering, waiting, terminate\}}.
    \item Start state: $S_0=active$.
    \item Input: $\Sigma=\{e_i\}_{i=1}^6$, per Table \ref{tab:mealy_inputs}.
    \item Output: $\Lambda=\{a_i\}_{i=1}^7$; actions $a_i$ correspond to data buffer decisions per Table~\ref{tab:mealy_outputs}.
    \item Transitions $T:S\times\Sigma \rightarrow S$ per Fig. \ref{fig:state_machine}.
    \item Output function $G:S\rightarrow\Lambda$:  mapping from states to outputs, per Table~\ref{tab:mealy_outputs}.
\end{itemize}

\subsection{Mealy Machine States}

\textit{Active}: The DMM is initialized in the active state with a precursor buffer $B_{pre}$ containing contextual data frames. Given input, the DMM transfers to the buffering or waiting state. Four inputs are possible for this state: $\{e_1, e_2, e_3, e_4\}$.

\textit{Buffering}: In the buffering state, a new data frame is stored in ``major" buffer $B_{maj}$. Data in $B_{maj}$ will eventually be used for LBO when the DMM  terminates. The DMM can transit to buffering, waiting or terminate states from the buffering state according to received inputs. All input are possible except $e_6$.

\textit{Waiting}: In this state, a new data frame is stored in a ``wait" buffer $B_{wait}$. $B_{wait}$ will be emptied when the DMM transits to buffering state or terminates. The machine can transit to the waiting, buffering or terminate state from waiting state based on inputs. All input are possible except $e_5$.

\textit{Terminate}: The DMM terminates once transits to the terminate state and the resulted $B_{maj}$ is sent to the following modules. A new DMM execution cycle will be initialized to track the next buffer. 

\subsection{Input alphabet:}

The input alphabet is generated based on data value, data similarity, major buffer size, and waiting time. 

Data value is estimated from \eqref{eq:eoi_value} and \eqref{eq:cutin_value_inv}. We set a threshold so that a frame $f_t$ with value $v_t>v(\varepsilon_1)$ indicates an EOI. Data similarity $\xi(f_t, B_k)$ is computed in \eqref{eq:similarity}. Note that $B_k$ is the major buffer $B_{maj}$ if it is not empty; otherwise, $B_k$ is the wait buffer $B_{wait}$. We set threshold $\xi_0$ so that  $\xi(f_t,B_k)>\xi_0$ indicates that frame $f_t$ and buffered data $B_k$ are from similar driving scenarios. The DMM thus appends $f_t$ to $B_k$ unless the buffer size limit is reached.

The major buffer size (number of frames) is represented as $|B_{maj}|$. We set a threshold so that the DMM state transits to \textit{terminate} when it reaches the largest allowed size $T_{maj}$ (event $e_5$). The waiting time is represented by the size of wait buffer $|B_{wait}|$. When DMM state is \textit{buffering}, $B_{wait}$ is empty so that waiting time is $0$. Each time the state transits to \textit{waiting}, waiting time will be incremented. We set a threshold so that the DMM state transits to \textit{terminate} when it has been waiting for more than $T_{wait}$ frames (event $e_6$). The elements in input alphabet $\Sigma$ are defined in Table \ref{tab:mealy_inputs}. 

\begin{table}[htbp]
    \centering
    \caption{DMM input alphabet, $\xi$ and $v$ are estimated similarity and value metrics.}
    \setlength{\extrarowheight}{5pt}
    \begin{tabular}{c|p{6cm}}
        \Xhline{3\arrayrulewidth}
        $\bm{\Sigma}$ & \textbf{Description} \\
        \hline
        $e_1$  & $\xi\leq\xi_0$ and $v\leq v(\varepsilon_1)$ and $|B_{wait}|< T_{wait}$ and $|B_{maj}|< T_{maj}$\\
        \hline
        $e_2$  & $\xi\leq \xi_0$ and $v>v(\varepsilon_1)$ and $|B_{wait}|< T_{wait}$ and $|B_{maj}|< T_{maj}$ \\
        \hline
        $e_3$  & $\xi>\xi_0$ and $v\leq v(\varepsilon_1)$ and $|B_{wait}|< T_{wait}$ and $|B_{maj}|< T_{maj}$ \\
        \hline
        $e_4$  & $\xi>\xi_0$ and $v>v(\varepsilon_1)$ and $|B_{wait}|< T_{wait}$ and $|B_{maj}|< T_{maj}$ \\
        \hline
        $e_5$  & $|B_{maj}|\geq T_{maj}$ \\
        \hline
        $e_6$  & $|B_{wait}|\geq T_{wait}$ \\
        \Xhline{3\arrayrulewidth}
    \end{tabular}
    \label{tab:mealy_inputs}
\end{table}

\begin{table}[htbp]
    \centering
    \caption{DMM output alphabet}
    \setlength{\extrarowheight}{5pt}
    \begin{tabular}{c|c|c|p{5cm}}
        \Xhline{3\arrayrulewidth}
        $\bm{S}$ & $\bm{\Sigma}$ & $\bm{\Lambda}$ & \textbf{Description} \\
        \hline
        \multirow{2}{*}{$s_1$} & $e_2$/$e_4$ & $a_1$ & Write from $B_{pre}$ to $B_{maj}$, empty $B_{pre}$ \\
        \cline{2-4}
        & $e_1$/$e_3$ & $a_2$ & Write from $B_{pre}$ to $B_{wait}$, empty $B_{pre}$ \\
        \Xhline{2\arrayrulewidth}
        \multirow{2}{*}{$s_2$} & $e_1$ & $a_5$ & Write new frame to $B_{wait}$.\\
        \cline{2-4}
        & $e_2$/$e_3$/$e_4$ & $a_3$ & Write new frame to $B_{maj}$ \\
        \cline{2-4}
        & $e_5$ & $a_6$ & Write last $L$ frames of $B_{maj}$ to $B_{pre}$, the rest of $B_{maj}$ is sent to LBO for long-term storage. \\
        \Xhline{2\arrayrulewidth}
        \multirow{2}{*}{$s_3$} & $e_1$/$e_3$ & $a_5$ & Write new frame to $B_{wait}$.\\
        \cline{2-4}
        & $e_2$/$e_4$ & $a_4$ & Write frames of $B_{wait}$ and the new frame to $B_{maj}$, empty $B_{wait}$\\
        \cline{2-4}
        & $e_6$ & $a_7$ & Write last $L$ frames of $B_{wait}$ to $B_{pre}$, and the rest of $B_{wait}$ to $B_{maj}$. $B_{maj}$ is sent to LBO for long-term storage. Empty $B_{wait}$. \\
        \Xhline{3\arrayrulewidth}
    \end{tabular}
    \label{tab:mealy_outputs}
\end{table}

\subsection{Output alphabet}
The output alphabet corresponds to buffer operations or actions given the current state and the input. 
Buffer operations include writing data to a buffer, writing data from one buffer to another, and emptying a buffer (see Table \ref{tab:mealy_outputs}). Typically a buffer will be emptied when its data is written to another buffer.

\begin{figure}[htbp]
    \centering
    \includegraphics[width=1.0\columnwidth]{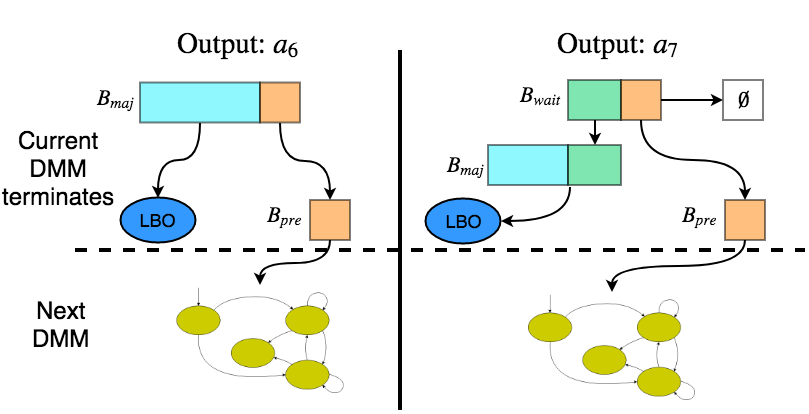}
    \caption{Buffer operation after DMM terminates.}
    \label{fig:output5_6}
\end{figure}

In the proposed DMM, $a_1$ to $a_5$ are outputs assigned during the buffer tracking process; these actions simply write to a buffer or empty a buffer. When the DMM terminates, either $a_6$ or $a_7$ is applied (Fig. \ref{fig:output5_6}). If the DMM transfers from the buffering state to the terminate state, $a_6$ executes. The last $L$ frames of $B_{maj}$ are used as $B_{pre}$ for the next buffer tracking cycle to provide contextual information. The previous frames are sent to LBO for data compression decision making. If the DMM transfers from the waiting state to the terminate state, $a_7$ executes. $B_{wait}$ is then divided into two partitions; the first (earliest) partition is appended to $B_{maj}$ while the most recent (latest) partition is used as $B_{pre}$ for the next DMM execution cycle. 

It is possible for $B_{maj}$ to overflow when populating it from $B_{wait}$, so the size of $B_{pre}$ is computed by:
\begin{equation}
    |B_{pre}| = \max(L, |B_{maj}| + |B_{wait}| - T_{maj})
\end{equation}
where $L$ is a user-specified minimum size of $B_{pre}$.

\section{Local Buffer Optimization and Long-term Storage Management}\label{sec:LBO}

This section specifies the LBO problem and proposes a method of decoupling LBO to facilitate real-time execution. LBO determines optimal compression quality of each frame in a buffer. A long-term storage management strategy is then introduced to deal with finite storage limits.

\subsection{LBO Formulation}

LBO is applied to each data buffer obtained from the DMM to determine the optimal compression quality for each frame of the buffer. This paper formulates LBO as a nonlinear programming (NLP) problem over design vector $D = [d_1, ..., d_{|B_k|}]$.
The objective function is based on three metrics: 1) Minimize total data storage cost; 2) Maximize total data value; 3) Maximize data recording decision continuity. These metrics are defined below.

\subsubsection{Storage cost (size)} 
For data buffer $B_k$, the total storage cost given decision vector $D$ is computed as

\begin{equation}\label{eq:memory_term}
    C_{B_k}(D) = \sum_{i=1}^{|B_k|} c_i \phi(d_i) \leq \sum_{i=1}^{|B_k|} c_i
\end{equation}
where $c_i$ is the storage cost of the $i$th frame in $B_k$, $\phi(d_i)$ is the mapping from compression quality to compression ratio, also called the quality-ratio curve. $\phi(d_i)$ monotonically increases over $d_i\in[0,1]$. 

The form of $\phi(d_i)$  depends on the compression algorithm and the data type. In this paper, we use $\phi(\cdot)$ of the JPEG compressor in \eqref{eq:quality_ratio},
\begin{equation}\label{eq:quality_ratio}
    \phi(d) = -a_1\log_2(1-a_2\,d)+a_3
\end{equation}
where $[a_1,a_2,a_3]$ is the parameter vector fit by compressing real-world driving videos. Readers are guided to~\cite{yao2018sbb} for further details.

\subsubsection{Data value term}
The total data value of $B_k$ given $D$ is computed in \eqref{eq:value_term}: 
\begin{equation}\label{eq:value_term}
    V_{B_k}(D)=\sum_{i=1}^{|B_k|} v_i d_i \leq \sum_{i=1}^{|B_k|} v_i
\end{equation}
Data value is presumed proportional to data compression quality in this work.
\subsubsection{Decision continuity term}
The decision continuity metric discourages abrupt changes in data frame decision value and is computed as the total change over all adjacent decisions in $D$. 

\begin{equation}\label{eq:continuity_term}
   W_{B_k}(D)=\sum_{i=2}^{|B_k|}(d_i-d_{i-1})^2
\end{equation}

This continuity term encourages storage of low-value frames when they are in proximity to high-value frames. Coupling is introduced between any two consecutive decisions to smooth the compression quality curve. 

\subsubsection{Objective function}
\ from \eqref{eq:memory_term}-\eqref{eq:continuity_term}: 
\begin{equation}\label{eq:lbo}
    \begin{split}
        \min_{D} & \quad\eta\, C_{B_k}(D) -\zeta\, V_{B_k}(D) + W_{B_k}(D)\\
        &\text{subject to } d_i\in[0,1],\quad \forall\, d_i\in D
    \end{split}
\end{equation}
Above, $\eta,\zeta\geq 0$ are weighting parameters that can be varied to examine solution sensitivity or represent user preferences. 

The optimization problem \eqref{eq:lbo} may not be easy to solve because the dimension $|B_k|$ is typically large. In what follows we introduce a simple but effective value filtering method so that the continuity term $W_{B_k}(D)$ can be dropped.  This results in a decoupled LBO problem where a unique minimizer exists and can be analytically solved.

\subsection{Event Value Filtering}
Estimating data value over sequential frames generates discrete-time value sequence $\{v_1,v_2,...,v_t,...\}$. This value sequence can have impulsive and step behaviors due to transient data events. The formulated LBO solves this problem by encouraging decision continuity $W_{B_k}(D)$ in \eqref{eq:continuity_term}. However, this term introduces coupling to LBO which results in a high-dimensional NLP.
As an alternative, we apply a data value filtering pre-processing step which suggests the $W_{B_t}(D)$ in \eqref{eq:lbo} can be eliminated. Data filtering is based on: 1) Assigning contextual frames of a high-value event high data value; 2) Preserving (not filtering out) impulsive events or short-term durational events of high value. 
The sequential data frame value signal is therefore filtered by Gaussian functions as shown in \eqref{eq:gaussian}.

\begin{equation}\label{eq:gaussian}
    \hat{v}_{t}= 
    \begin{cases}
        \max(v_{t},v_{T_0}e^{-(\frac{t-T_0}{\sigma_f})^2}) & \text{if $t<T_0$}\\
        \max(v_t,v_{T_1}e^{-(\frac{t-T_1}{\sigma_f})^2}) & \text{if $t>T_1$}\\
        v_t & \text{otherwise}
    \end{cases}
\end{equation}
where $T_0$ and $T_1$ are event start and end time, respectively, and $\sigma_f$ is data value deviation that controls Gaussian curve width. 

\subsection{Decoupled LBO}
The proposed data value filtering scheme can serve to decouple consideration of data continuity from LBO computation.  We can drop the continuity term in \eqref{eq:lbo} to  
decouple the overall solution into a series of one-dimensional frame optimization problems per \eqref{eq:single_opt}.  With this strategy, LBO can optimize the data value specification for each frame independent of all other frames. Given constant $\frac{\zeta}{\eta}$, each LBO decision is determined by a data frame's  value and size.
\begin{equation}\label{eq:single_opt}
    \begin{split}
        \min_{d_i}  c_i\phi(d_i)-\frac{\zeta}{\eta}\, \hat{v}_i\, d_i,\\
        \text{subject to } d_i\in[0,1].
    \end{split}
\end{equation}

It has been shown in~\cite{yao2018sbb} that the decoupled LBO performs similar to coupled LBO with much less computation time. 

\subsection{Prioritized Data Recording in Long-term Storage}

Once SBB storage limit $M$ is reached and an optimal decision vector for a new buffer is computed, either old buffer(s) or the new buffer must be discarded. This paper proposes storing buffers over a long-term as a priority queue (heap) so that those with lower values are discarded preferentially. The heap is constructed based on buffer value, and the lower-value buffers are discarded until heap size is less than $M$. 

\subsubsection{Stored Data Buffer Value and Size}
The total value of the $k$th buffer is calculated as:
\begin{equation}\label{eq:buf_value}
    V_{B_k}^* = (1+\lambda)^k\max_i(v_i\cdot d_i)
\end{equation}
where $v_i$ is the $i$th data frame value, and $1+\lambda$ with $0<\lambda\ll1$ is an aging factor that amplifies the value of the newest data buffer. 

\subsubsection{Prioritized Data Recording}
Each recorded buffer $\hat{B}_k$ contains the compressed data for all $f_t\in B_k$. The buffer storage cost is $C_{B_k}$ as in \eqref{eq:memory_term} and the buffer value is $V_{B_k}^*$. When a buffer is to be removed, data included in the buffer and their indices can be rapidly located and pruned; note that pruned indices will not be reused in other buffers. In this paper, a binary $min-heap$ queue~\cite{thomascormen2009} is constructed to store buffers based on $V_{B_k}^*$. Algorithm \ref{alg:heap} describes the prioritization sequence.

\begin{algorithm}
    \caption{Priority Queue Logic}
    \label{alg:heap}
    \SetKwInOut{Input}{Input}
    \SetKwInOut{Output}{Output}
    \Input{New buffer $B_k$, heap $HQ$, heap size $C_{HQ}$, maximum available storage $M$.}
    \Output{Updated heap $HQ$}
    HQ.push($B_k$)\\
    $C_{HQ} = C_{HQ} + C_{B_k}$\\
    \While{$C_{HQ} > M$}{
        $\widehat{B}$ = HQ.pop() \tcp{pop the buffer with the smallest $V_{\widehat{B}}$}
        $C_{HQ} = C_{HQ} - C_{\widehat{B}}$
    }
\end{algorithm}

\section{Case Studies and Results}\label{sec:case}

This section presents a case study using the pre-defined EOIs from Section~\ref{sec:value_similarity} and long-term traffic data generated from a simulator. Two case studies are presented analyzing coupled and decoupled LBO parameter selection, respectively. A comparison between prioritized data recording and FIFO recording with different storage limitations is also provided. Last, we examine the reproducibility of deep learning results on images compressed by the SBB to demonstrate its utility.

\subsection{Simulation Environment}

A simulator is used to generate three-lane highway traffic trajectories for this case study. The simulator is developed based on a game theoretic traffic model and is capable of representing heterogeneous and interactive multi-vehicle traffic scenarios. More details of the simulator can be found in~\cite{Li2017game}. We feed data generated from the simulator into our SBB. This simulator is utilized because it covers a large range of traffic scenarios over a short period of running time. Note that the proposed approach can be applied to other datasets also.

In this case study, we define one host car and 15 participant cars so that the EOIs are not too frequent or rare. The frame rate is $10 Hz$ and the trajectory length is $600$ seconds.

\xhdr{Training data.} We generated $10,000$ Monte Carlo trajectories with randomly initialized car locations and velocities to estimate the data value metrics. Each trajectory terminates when the set length is reached or when the host car crashes with a participant car. Driving data of the host car and all participant cars are collected. The accumulated driving time of all cars is about $15,041$ hours and the total mileage is about $0.68$ million kilometers.

From the 10,000 MC trajectories, there are $29,953,405$ cut-in events captured among all cars as well as $23,000,206$ hard braking events and $1,024,611$ conflicts. Since the simulation is restarted if one host car crash is detected, we compute probability of crash using only host car crashes; $4,799$ crashes are obtained. Likelihood of events are computed from \eqref{eq:prior} and listed in Table \ref{tab:prob_value}. For crash probability, the denominator is the number of frames for the host car only.

\begin{equation}\label{eq:prior}
    \Pr(\varepsilon_j) = \frac{\text{\# of frames with $\varepsilon_j$ detected}}{\text{total \# of MC trajectory frames}}
\end{equation}

\begin{table}[htbp]
    \centering
    \setlength{\extrarowheight}{5pt}
    \caption{Probability and estimated value metrics of normal frames and EOIs. $cutin_1$ and $cutin_2$ have ranges $R=100\,m$ and $30\,m$, respectively.}
    \scriptsize{
    \begin{tabular}{c|ccccc}
        \toprule
        \textbf{Events} & $normal$ & $cutin_1$ & $cutin_2$ & $hardbraking$ & $conflict$ \\
        \midrule
        \textbf{Prob.} &0.92 & 0.045 & 0.010 & 0.035 & 0.0015 \\
        \midrule
        \textbf{Value} & 0.009 & 0.34 & 0.53 &  0.37 & 0.72 \\
        \bottomrule
    \end{tabular}
    }
    \label{tab:prob_value}
\end{table}

\xhdr{Test data.} We simulated a single long-term test trajectory to evaluate SBB performance. This long-term trajectory consisted of $115,615$ frames (around $3$ hours $12$ minutes) and terminated with a crash. Statistics on test data are shown in the first and second rows of Table~\ref{tab:sbb_stat}.
We evaluated the SBB with coupled and decoupled LBO over this trajectory, then compared prioritized data recording with a conventional FIFO queuing model. We also considered the reproducibility of deep learning results on SBB compressed data for object detection and semantic segmentation models.

\xhdr{Metrics.} We define three metrics, average value per frame ($aVPF$), average memory per frame ($aMPF$) and value per memory ($VPM$), to guide parameter selection for LBO: 

\begin{subequations}\label{eq:metrics}
    \begin{align}
        aVPF & = \frac{\sum_i{\hat{v}_i\,d_i}}{N} \\
        aMPF & = \frac{\sum_i{c_i\,\phi(d_i)}}{N} \\
        VPM & = \frac{aVPF}{aMPF}
    \end{align}
\end{subequations}

\subsection{Case Study 1: Coupled LBO and Parameter Selection}\label{sec:case_study_1}

\begin{figure*}[htbp]
    \centering
    \begin{subfigure}[t]{0.32\textwidth}
        \centering
        \includegraphics[width = 1\columnwidth]{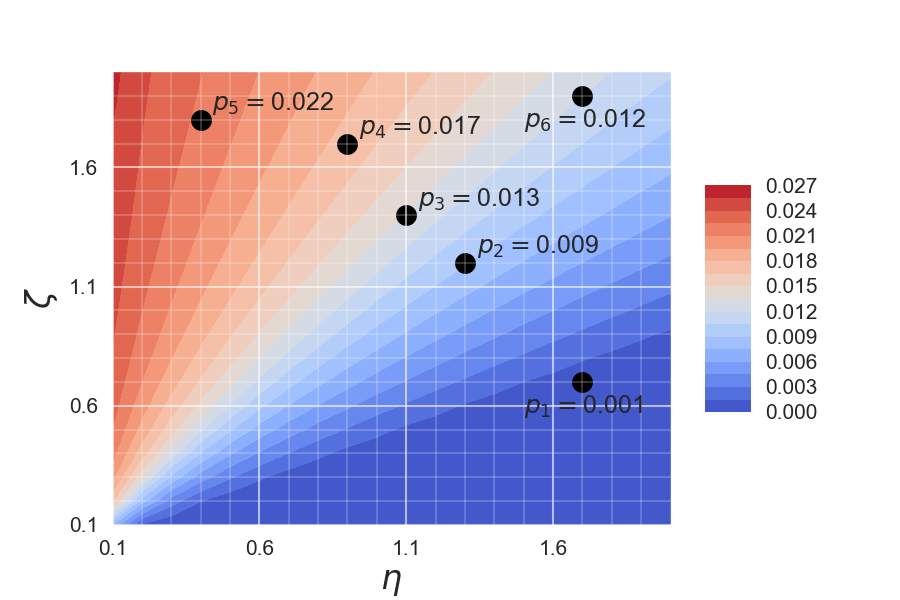}
        \caption{Average value per frame ($aVPF$) with different $\eta$ and $\zeta$}
        \label{fig:value3D_cntr}
    \end{subfigure}
    ~
    \begin{subfigure}[t]{0.32\textwidth}
        \centering
        \includegraphics[width = 1\columnwidth]{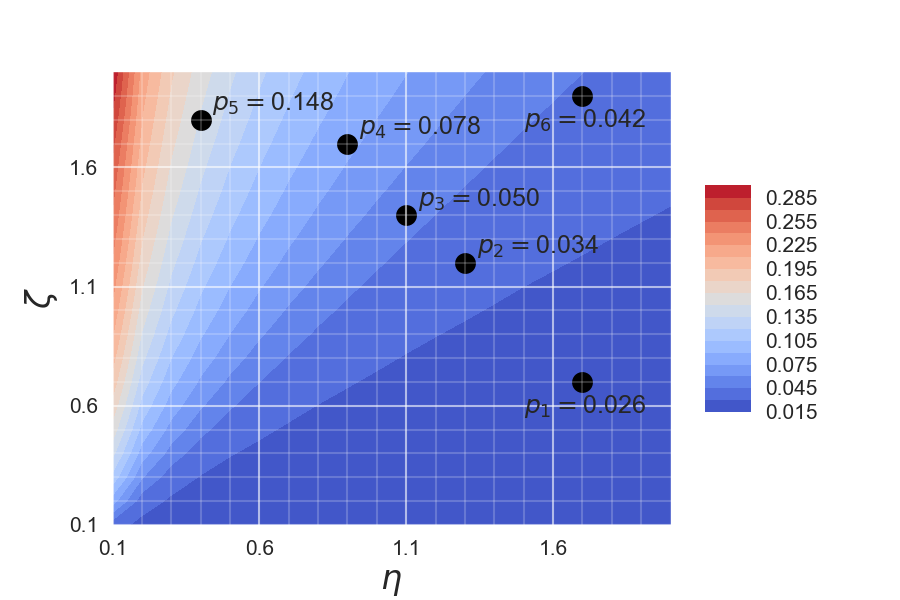}
        \caption{Average memory per frame ($aMPF$) with different $\eta$ and $\zeta$}
        \label{fig:memory3D_cntr}
    \end{subfigure}
    ~
    \begin{subfigure}[t]{0.32\textwidth}
        \centering
        \includegraphics[width = 1\columnwidth]{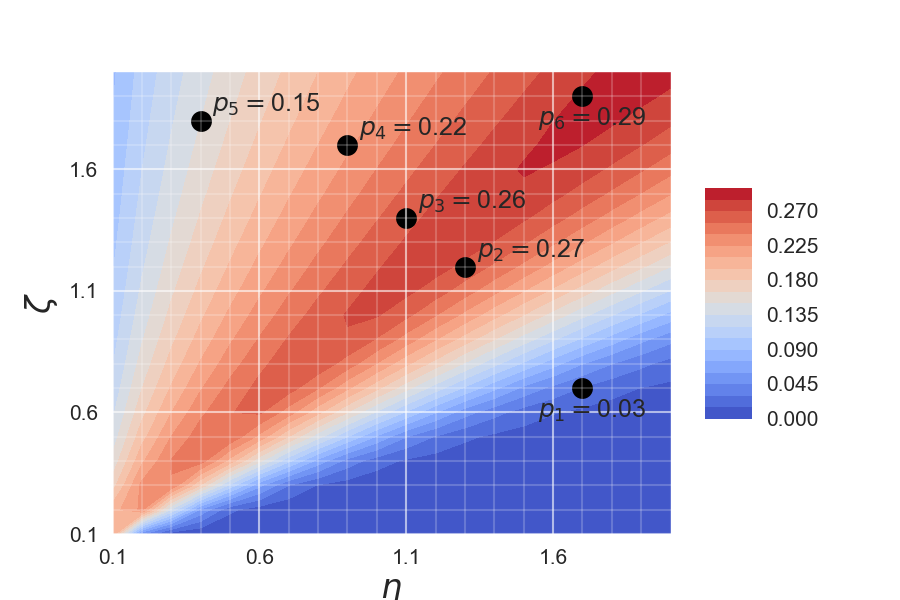}
        \caption{Value per memory ($VPM$) with different $\eta$ and $\zeta$}
        \label{fig:vpm3D_cntr}
    \end{subfigure}
    \caption{Sensitivity analysis of the coupled LBO method.}
    \label{fig:coupled_vpm_contour}
\end{figure*}

We first applied the proposed SBB with coupled LBO to the testing data. The sensitivity of coupled LBO to weighting parameter $\eta$ and $\zeta$ is investigated. Figs. \ref{fig:value3D_cntr} to \ref{fig:vpm3D_cntr} show the contours of $aVPF$, $aMPF$ and $VPM$ with $\eta\in\{0.1,0.2,...,2.0\}$ and $\zeta\in\{0.1,0.2,...,2.0\}$. Other parameters used throughout the paper include:  $T_{maj}=600$, $T_{wait}=30$, $L=20$, $\sigma_f=10$, $\xi_0=0.5$.

Generally, the SBB performance is best when selecting parameters that result in high $VPM$ value. However, the $VPM$ can be large as long as the $aMPF$ is small enough, in which case all data are highly compressed. Therefore, users might also have a minimum expectation regarding $aMPF$ (or $aVPF$) of recorded SBB data. Thus, instead of simply selecting $\eta$ and $\zeta$ corresponding to the highest $VPM$, the trade-off between $VPM$ and $aVPF$ must be considered.
Six example parameter selections ($p_1$ to $p_6$) are shown in Fig.~\ref{fig:coupled_vpm_contour} and their corresponding metric values are presented. We select $\eta=0.9$ and $\zeta=1.7$ for the following experiment since it results in relatively high $VPM$ and $aVPF$. 

Computational time required to solve the coupled LBO depends on  data buffer length and frame values.  The buffer length obtained by the DMM is from $3$ seconds to $60$ seconds with average $7.5$ seconds, while the LBO solving time varies from $0.003$ seconds to $24.6$ seconds with average $0.9$ seconds. In conclusion, solving a coupled LBO is time-consuming with a large buffer containing highly variable data values, motivating application of decoupled LBO in a deployed SBB.
In this study we ran sequential least squares programming (SLSQP) solver provided by a Scipy (Python) optimization package on a machine with 16GB RAM and an Intel Xeon(R) CPU E3-1240 v5 @ 3.50GHZ*8.

\subsection{Case Study 2: Decoupled LBO and Parameter Selection}

The continuity term in the decoupled LBO is dropped, 
resulting in \eqref{eq:decoupled_obj_func}. Parameters $[a_1,a_2,a_3]$ are obtained from quality-ratio curve fitting in~\cite{yao2018sbb}.

\begin{equation}\label{eq:decoupled_obj_func}
    \begin{split}
    F(d_i) & = \phi(d_i)-\frac{\zeta}{\eta}\, \hat{v}_i\, d_i\\
           & = -a_1\log_{2}(1-a_2d_i)+a_3-\frac{\zeta}{\eta}\, \hat{v}\, d_i
    \end{split}
\end{equation}
The resulting function is convex and derivable in $d_i\in[0,1]$. Therefore a unique minimum solution can be analytically computed by finding the zero-derivative solution. 
The optimal decision is then given by:
\begin{equation}
    d_i^*=\max(0,\, -\frac{a_1}{\log_{2}2\hat{v}_i}(\frac{\zeta}{\eta})^{-1} + \frac{1}{a_2})
\end{equation}

\begin{figure*}[htbp]
    \centering
    \begin{subfigure}[t]{0.3\textwidth}
        \centering
        \includegraphics[width = 1\columnwidth]{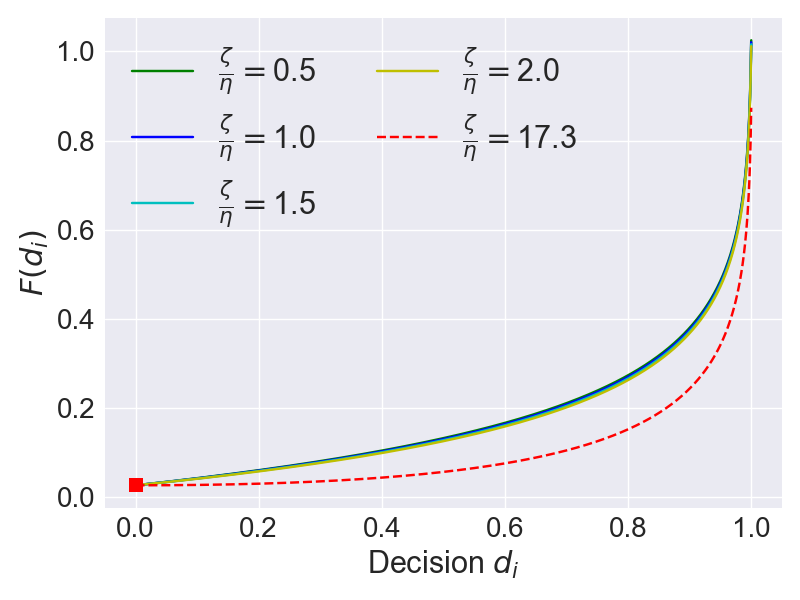}
        \caption{normal}
        \label{fig:eta1}
    \end{subfigure}
    ~
    \begin{subfigure}[t]{0.3\textwidth}
        \centering
        \includegraphics[width = 1\columnwidth]{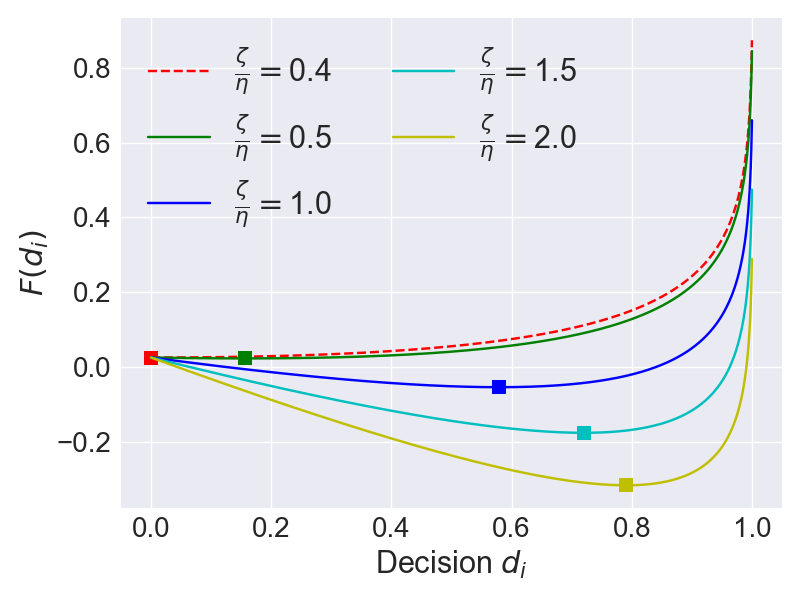}
        \caption{hard braking}
        \label{fig:eta5}
    \end{subfigure}
    ~
    \begin{subfigure}[t]{0.3\textwidth}
        \centering
        \includegraphics[width = 1\columnwidth]{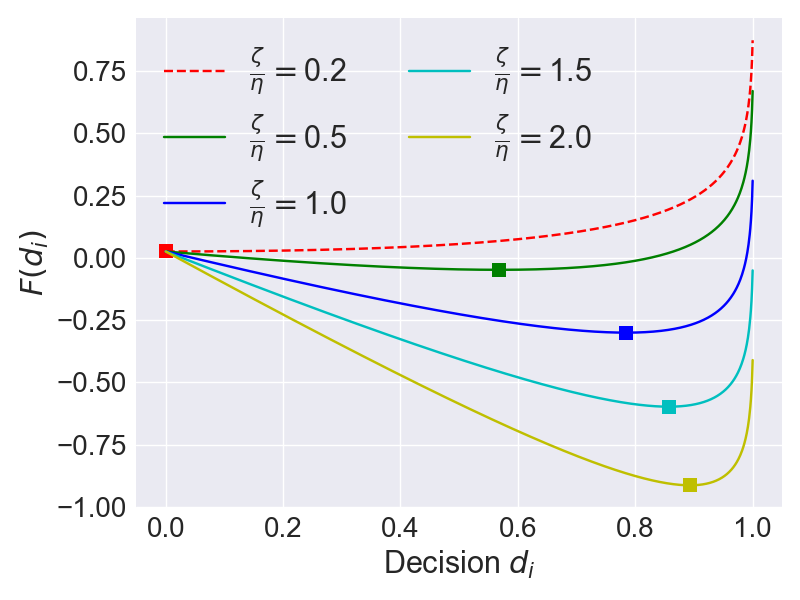}
        \caption{conflict}
        \label{fig:eta01}
    \end{subfigure}
    
    \caption{Objective function of decoupled LBO for each event. Square blocks indicate optimal solutions. Each red dotted curve indicates the boundary value of parameter for such an event. }
    \label{fig:decoupled_obj_func}
\end{figure*}

Fig. \ref{fig:decoupled_obj_func} shows the objective functions of four constant-value events with different $\frac{\zeta}{\eta}$. The $cutin$ event is not shown since its value is a function of observed $cutin$ range. This ratio must be selected in an interval such that all EOIs can be recorded with a high quality while $normal$ frames are compressed with low quality.

Given the fitted parameters, if
\begin{equation}\label{eq:boundary_param}
    \frac{\text{d}F(d_i)}{\text{d}d_i}\Big|_{d_i=0} \geq 0
\end{equation}
then $F(d_i)$ is monotonically increasing and the minimum is $d_i^*=0$. We define  boundary parameter $\frac{\zeta}{\eta} = \frac{a_1a_2}{(\log2\hat{v}_i)}$ by solving the equality condition in \eqref{eq:boundary_param}. There is one boundary parameter for each EOI. Corresponding objective functions are shown by red dotted curves in Fig. \ref{fig:decoupled_obj_func}. Parameter selection must have $\frac{\zeta}{\eta} > \frac{a_1a_2}{(\log2\hat{v})}$ for all EOIs. In this case study, to guarantee that the $hard braking$ event is recorded at high quality, we select $\frac{\zeta}{\eta}>0.4$. If $\frac{\zeta}{\eta}<0.2$, none of these EOIs are recorded. To avoid recording normal data with high quality, $\frac{\zeta}{\eta}<17.3$ should be enforced. 

\begin{figure}[htbp]
    \centering
    \includegraphics[width=1.0\columnwidth]{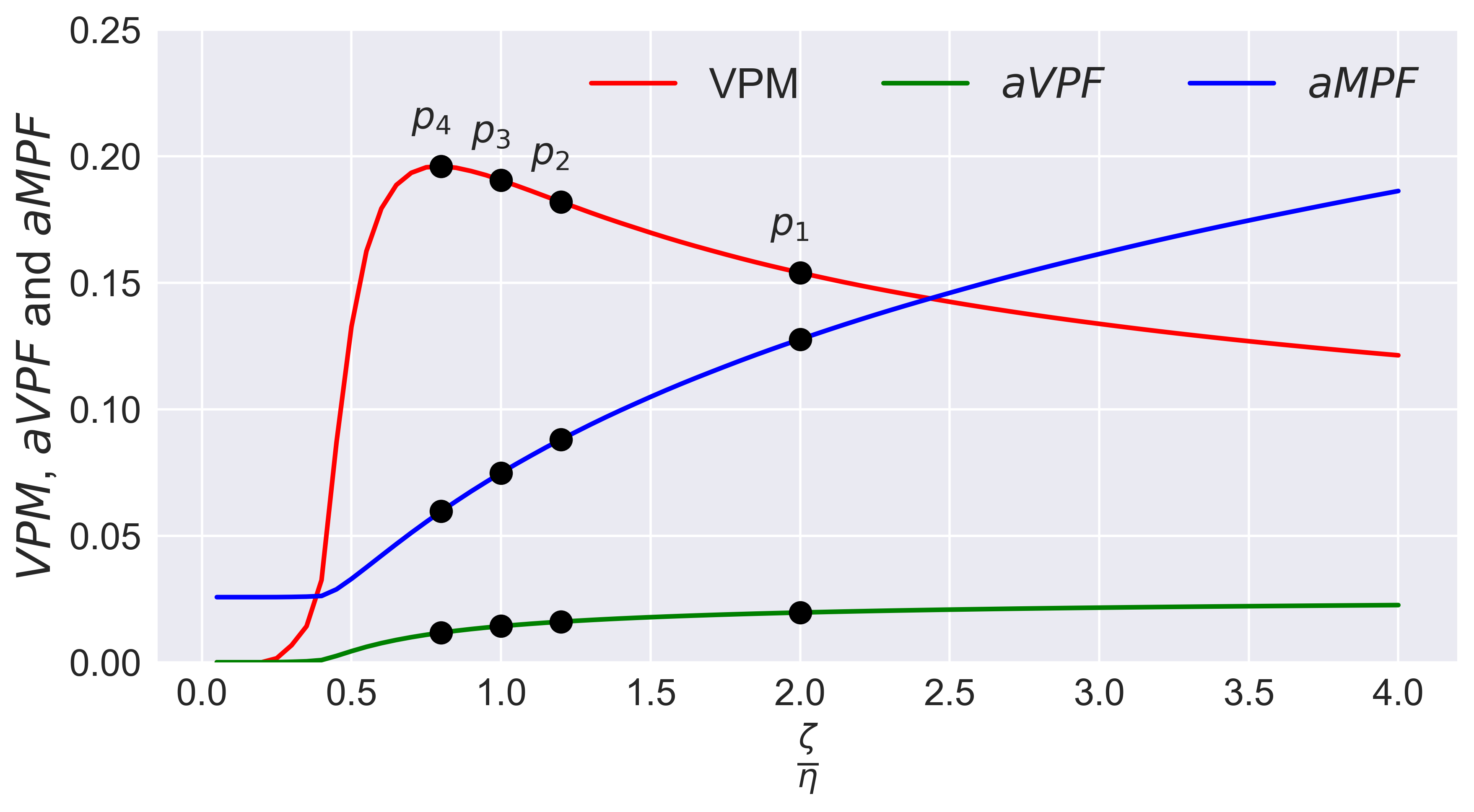}
    \caption{Value per frame with different weighting parameter ratios for the decoupled LBO method. Black points are examples of different parameter selections, similar to Fig.~\ref{fig:coupled_vpm_contour}.}
    \label{fig:zeta_eta_ratio}
\end{figure}

In Fig. \ref{fig:zeta_eta_ratio}, $VPM$ decreases significantly when $\frac{\zeta}{\eta}>0.7$. However, $aVPF$ and $aMPF$ values are extremely small when $\frac{\zeta}{\eta} < 0.7$, indicating that most data are highly compressed. This is consistent with observations from the coupled LBO parameter selection study. 
Therefore we recommend $\frac{\zeta}{\eta} > 0.7$ but not too large to realize a reasonable trade-off between $VPM$ and $aVPF$. 

\subsection{Results and Analysis} \label{sec:result_analysis}

Below we present SBB results using the coupled LBO method with parameters selected per Section~\ref{sec:case_study_1}.

\subsubsection{SBB data recording statistics}

Statistics of SBB data recording on the test data are presented in Table \ref{tab:sbb_stat}. The third and fourth rows show the average and standard deviation of SBB compression decision (quality) of each event. It can be seen that the qualities on EOIs are much higher and more stable compared to the quality of $normal$ frames, indicating that the SBB places high emphasis on all EOI frames but treats $normal$ frames differently based on how close each frame is to an EOI. Generally, the SBB maintains high quality for EOIs while compressing normal data frames to save memory.

\begin{table}[ht]
    \centering
    \caption{Raw test data and SBB data compression statistics.}
    \setlength{\extrarowheight}{5pt}
    \scriptsize{
    \begin{tabular}{c|c|cccc}
        \toprule
        & & $normal$ & $cutin$ & \makecell{$hard$\\$braking$} & $conflict$ \\
        \midrule
        \multirow{2}{*}{Raw data}& \textbf{\# of frames} & 106230 & 2955 & 5865 & 541 \\
        & \textbf{\makecell{Size (MB)}} & 19639.16 & 547.18 & 1082.15 & 101.45 \\
        \midrule
        \multirow{3}{*}{SBB data}& \textbf{Avg.} $d_i$ & 0.44 & 0.76 & 0.79 & 0.88 \\
        & \textbf{Std.} $d_i$ & 0.36 & 0.019 & 0.012 & 0.005 \\
        & \textbf{\makecell{Size (MB)}} & 1511.30 & 93.02 & 151.63 & 22.70 \\
        \bottomrule
    \end{tabular}
    }
    \label{tab:sbb_stat}
\end{table}

The last row of Table \ref{tab:sbb_stat} summarizes SBB memory requirements for each event type to support comparison of raw data and SBB storage requirements over normal and EOI datasets. Most of the recorded driving data frames are $normal$ since the SBB records all frames unless finite storage is reached and $normal$ frames are dominant in the test data set. Some $normal$ frames are also contextual frames of EOIs which have higher value according to the filter and therefore are compressed with high quality. Table \ref{tab:context_percentage} indicates the percentage (pct.), quantity, and storage cost of contextual frames as a fraction of all normal frames acquired from simulations. The contextual frames here are defined as frames within a specified range (in number of frames) from an EOI.

\begin{table}[htbp]
    \centering
    \caption{Statistics on recorded normal frames. The context range is in number of frames with frame rate $10 Hz$.}
    \setlength{\extrarowheight}{5pt}
    \begin{tabular}{c|cccc}
        \Xhline{3\arrayrulewidth}
        \textbf{Context range} & $\pm 5$ & $\pm 10$ & $\pm 15$ & $\pm 20$ \\
        \midrule
        \textbf{Quantity pct.} & $20.27\%$ & $34.99\%$ & $46.19\%$ & $54.80\%$ \\
        \textbf{Storage cost pct.} & $43.29\%$ &  $66.87\%$ & $79.46\%$ & $83.09\%$\\
        \Xhline{3\arrayrulewidth}
    \end{tabular}
    \label{tab:context_percentage}
\end{table}

For example, we found that $20.27\%$ of the $normal$ frames are located within $\pm 5$ frames of an EOI, and their storage cost is about $43.29\%$ of the storage cost of all $normal$ frames. This indicates that the memory is efficiently utilized to record contextual frames. $83.09\%$ of the storage cost of $normal$ frames are within $\pm 20$ frames of an EOI.

\subsubsection{Prioritized data recording with long-term storage limits}

We apply different storage limitations ($M$) to the SBB.  Recorded event counts are summarized in Table \ref{tab:heapq}. With $M=1500\,MB$, more $cutin$ ($\sim200$), $hardbraking$($\sim500$), and $conflict$($\sim90$) events are missed by the FIFO model compare to the prioritized model, while the SBB reserves more storage for EOIs by discarding more $normal$ frames. With $M=500\,MB$, the SBB model saves $40\%$ of the $cutin$ events and all $conflict$ events, while the FIFO model records $10,072$ more $normal$ frames and missed $79\%$ of the $conflict$ frames. 
Note that less $hardbraking$ frames are recorded by the SBB with $M=500\,MB$ because their values are determined to be lower than for some $cutin$s. These differences show the prioritized data recording scheme is able to record valuable data happening in the early phase of the trajectory that would be discarded by a conventional FIFO (circular buffer) data recorder.
\begin{table}[htbp]
    \centering
    \setlength{\extrarowheight}{5pt}
    \caption{Prioritized (SBB) and FIFO data recording comparison with storage limit $M$. Both schemes use LBO to guide data buffer JPEG compression.}
    \begin{tabular}{c|c|cccc}
        \Xhline{3\arrayrulewidth}
                          $M$ &  & $normal$ & $cutin$ & $hard braking$ & $conflict$ \\
        \midrule
        \multirow{2}{*}{$1500MB$} & \textbf{Ours} & $74271$ & $2725$ & $5547$ & $541$ \\
        & \textbf{FIFO} & $90953$ & $2530$ & $5036$ & $456$  \\
        \midrule
        \multirow{2}{*}{$500MB$} & \textbf{Ours} & $21135$ & $1182$ & $1599$ & $541$ \\
        & \textbf{FIFO} & $31207$ & $738$ & $1719$ & $113$  \\
        
        \Xhline{3\arrayrulewidth}
    \end{tabular}
    \label{tab:heapq}
\end{table}

\subsection{Performance of SBB Data on Deep Learning Model}

To evaluate SBB compression with respect to AV percept reproducibility, we applied object detection and semantic segmentation models on data recorded with SBB compression and two comparative baselines. For object detection, we apply Mask-RCNN~\cite{He2017} pre-trained on COCO dataset. The bounding box average precision $AP^{bb}$~\cite{He2017} is computed as the evaluation metric. For semantic segmentation tasks, we use DeepLabV2~\cite{CP2016Deeplab} pre-trained on the crowdflower dataset. We evaluate the performance using pixel intersection over union (IOU) as shown in \eqref{eq:piou}. 
\begin{equation}\label{eq:piou}
    PIOU = \frac{\sum^{W,H}_{i=0,j=0} \mathds{1}_(\hat{p}_{i,j},p_{i,j})}{W\cdot H}
\end{equation}
where $\hat{p}_{i,j}$ and $p_{i,j}$ are predicted and ground truth classes of pixel $(i,j)$, respectively, and $\mathds{1}_{i,j}$ is an indicator function returning $1$ if $\hat{p}_{i,j}=p_{i,j}$.

The \textbf{SBB} compresses images using the method from Section~\ref{sec:result_analysis}  resulting in $1778.91\,MB$ of stored data; \textbf{Baseline1} compresses all images with $0.1$ quality resulting in $1433.6\,MB$ of data; \textbf{Baseline2} compresses all images with $0.5$ quality, resulting in $2355.2\,MB$ of data.

\begin{figure}[ht]
    \centering
    \includegraphics[width=0.9\columnwidth]{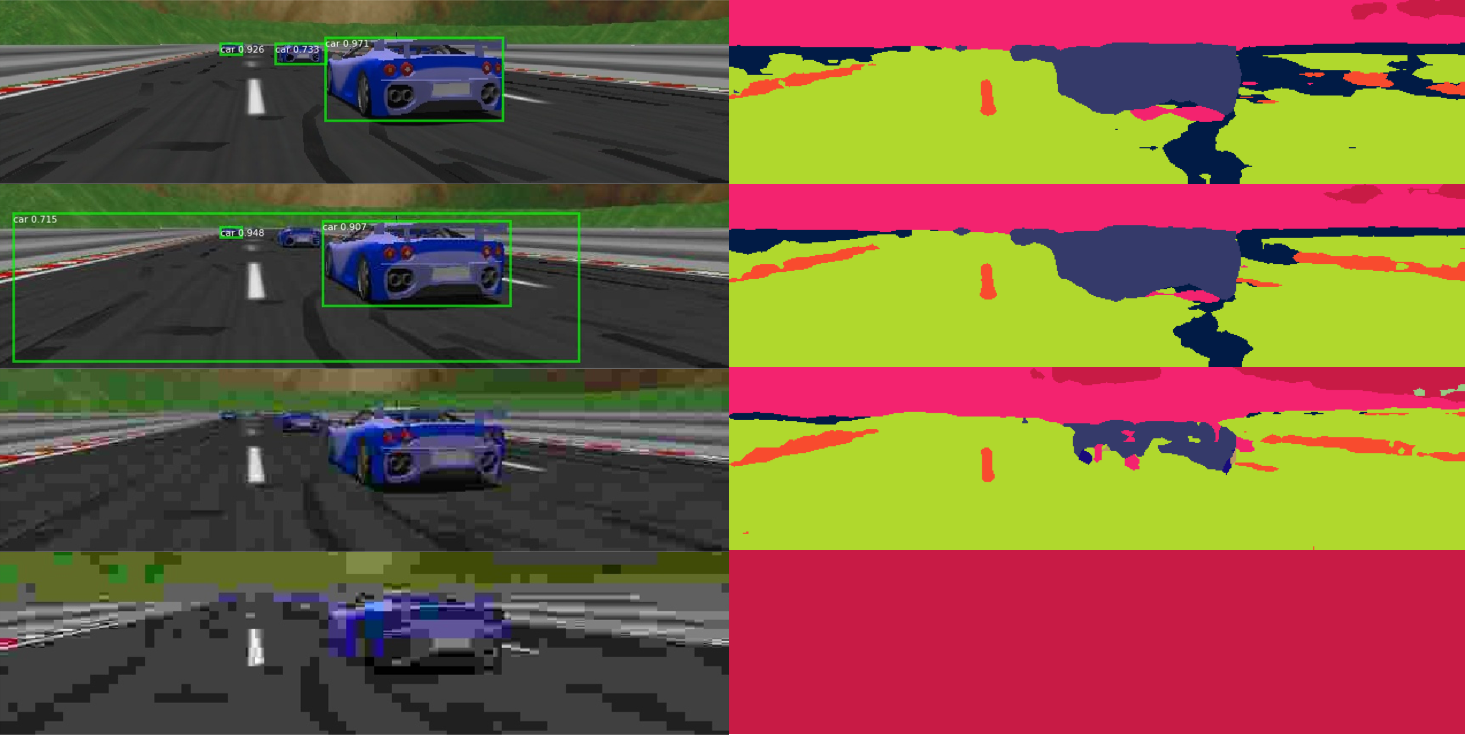}
    \caption{Object detection (left) and semantic segmentation (right) on an TORCS image compressed with 1, 0.5, 0.1 and 0.01 quality (from top to bottom).}
    \label{fig:torcs_result}
\end{figure}

Image data is obtained from a high-fidelity simulator called TORCS  which generates first-person driving video. The simulator~\cite{Li2017game} used to generate the traffic data has been integrated with TORCS to assure experimental data is consistent\cite{su2018torcs}. TORCS-generated images are referred to as ``raw data" in this study and SBB compressed images are ``SBB data". Fig.~\ref{fig:torcs_result} shows an example result with different compression qualities, consistent with Fig.~\ref{fig:compress_intro}.

The metrics values of Mask-RCNN and DeeplabV2 on different data are presented in Table \ref{tab:eval_dl}. We assume object detection and semantic segmentation results on raw data are ground truth and report the metrics with SBB and two baseline datasets. We separately show the results on $normal$ frames and frames contain 3 EOIs and ignore crash frames since no compression is applied with SBB. Both Mask-RCNN and DeeplabV2 achieve better performance on EOI frames in SBB data compared to baseline data, indicating that the EOIs saved by the SBB is more reproducible in these two specific tasks. The performance on $normal$ images recorded by the SBB is worse than on the images from baselines, consistent with the SBB design goal to highly compress normal data to save memory for EOIs. The SBB is able to record more reproducible but less memory-consuming  data compared to  compressing data with a preset constant compression ratio. 

Another interesting observation is that DeeplabV2 performance is more robust to JPEG compression compared to Mask-RCNN. One possible reason is that with JPEG compression, the object feature used for detection is destroyed while the whole image feature is better maintained, which makes it harder to distinguish an object from background.

\begin{table}[h]
    \centering
    \caption{Mask-RCNN $AP^{bb}$ and DeeplabV2 $PIOU$ on two baseline data and SBB data.}
    \setlength{\extrarowheight}{5pt}
    \scriptsize{
    \begin{tabular}{c|c|cccc}
        \toprule
        & \textbf{Data} & $normal$ & $cutin$ & \makecell{$hard$\\$braking$} & $conflict$ \\
        \midrule
        \multirow{3}{*}{\makecell{Mask\\RCNN}}& Baseline1  & 0.32 & 0.32 & 0.38 & 0.28 \\
        & Baseline2 & 0.68 & 0.68 & 0.74 & 0.67\\
        & SBB & 0.47 & 0.74 & 0.80 & 0.80 \\
        \midrule
        \multirow{3}{*}{DeeplabV2}& Baseline1 & 0.78 & 0.80 & 0.79 & 0.79 \\
        & Baseline2 & 0.88 & 0.90 & 0.90 & 0.91 \\
        & SBB & 0.58 & 0.92 & 0.92 & 0.95\\
        \bottomrule
    \end{tabular}
    }
    \label{tab:eval_dl}
\end{table}

\section{conclusion}\label{sec:conclusion}

This paper has presented a Smart Black Box (SBB) architecture that makes compression and storage prioritization decisions with a two-stage process. The SBB first caches raw data in a short-term buffer and determines compression factor based on data value and size. Data value is computed based on its novelty and the presence of temporally-proximal high-value data frames. The short-term buffers are managed by a deterministic Mealy machine so that high-value data or similar data are buffered together. For long-term data collection given finite onboard storage, the SBB discards the lowest-value data regardless of age. A simulation case study generates driving trajectories and first-person view images containing four predefined EOIs. We show that the local buffer optimization strategy enables the SBB to record reproducible but less memory-consuming data. 

The proposed SBB makes decisions by optimizing compression over a static combination of data value, data size and continuity metrics. Future work should extend the SBB to combine static metrics with real-time (dynamic) metrics including available storage, observed driving and traffic risk, and observed frequency (novelty) of each event type. 

\bibliographystyle{IEEEtran}
\bibliography{references}

\vskip 0pt plus -1fil
\begin{IEEEbiography}[{\includegraphics[width=1in,height=1.25in,clip,keepaspectratio]{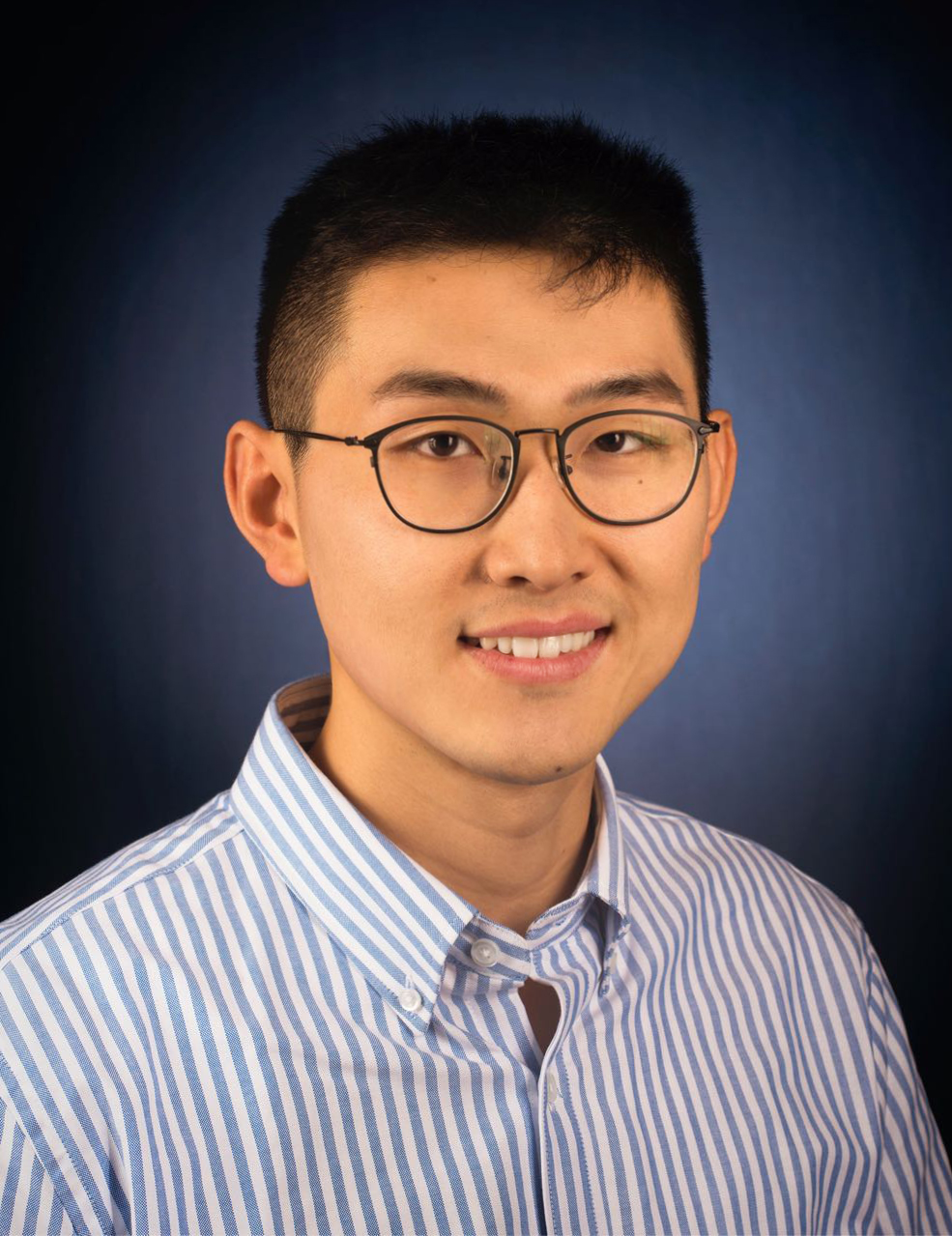}}]
{\textbf{Yu Yao}} received the B.Eng. degree in Aerospace Engineering from Beijing Institute of Technology in 2015 and M.S. degree in Robotics from the University of Michigan in 2017. He is currently a PhD candidate at the University of Michigan Robotics Institute. His research interests includes anomaly detection, action recognition/prediction, scene understanding and their applications to autonomous vehicles and intelligent transportation systems.
\end{IEEEbiography}
\vskip 0pt plus -1fil

\begin{IEEEbiography}[{\includegraphics[width=1in,height=1.25in,clip,keepaspectratio]{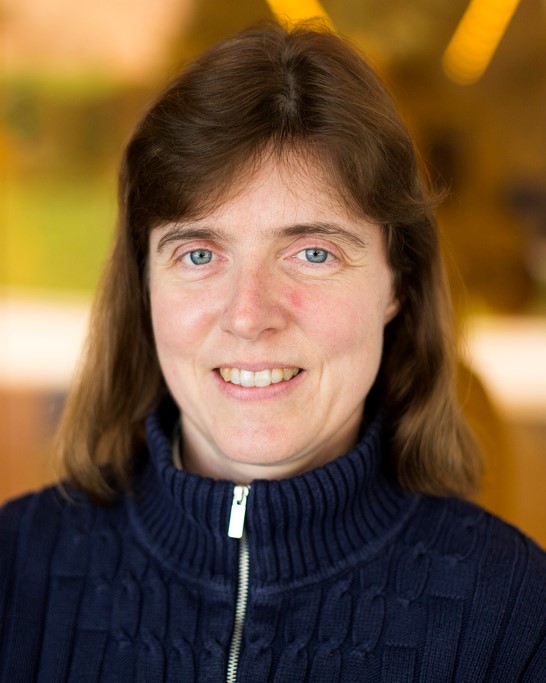}}]
{\textbf{Dr. Ella Atkins}} is a Professor of Aerospace Engineering at the University of Michigan, where she directs the Autonomous Aerospace Systems Lab and is Associate Director of Graduate Programs for the Robotics Institute.  Dr. Atkins holds B.S. and M.S. degrees in Aeronautics and Astronautics from MIT and M.S. and Ph.D. degrees in Computer Science and Engineering from the University of Michigan.  She is Editor-in-Chief of the AIAA Journal of Aerospace Information Systems (JAIS) and past-chair of the AIAA Intelligent Systems Technical Committee. She has served on the National Academy's Aeronautics and Space Engineering Board, and pursues research in Aerospace and robotic system contingency planning, autonomy, and safety.
\end{IEEEbiography}

\end{document}